\documentclass[prd,twocolumn,preprintnumbers,amsmath,amssymb,nofootinbib,english,showpacs]{revtex4}

\usepackage{mathrsfs}
\usepackage{graphics}
\usepackage{epsfig}
 \usepackage{graphicx}
\usepackage{bm}
\usepackage{hyperref}
\usepackage[usenames]{color}
\usepackage{url}

\hypersetup{
    colorlinks=true,
    linkcolor=red,
    citecolor=red,
}

\newcommand{\dw}{\textcolor{black}}
\newcommand{\tc}{\textcolor{black}}
\newcommand{\mv}{\textcolor{black}}
\newcommand{\mvs}{\textcolor{black}}
\newcommand{\remove}[1]{}

\def\ie{{\frenchspacing\it i.e.}}
\def\eg{{\frenchspacing\it e.g.}}

\begin{document}

\title{\tc{Post-$Planck$ constraints on interacting vacuum energy}}

\author{Yuting Wang$^{1,}$$^{2}$, David Wands$^{2}$, Gong-Bo Zhao$^{1,}$$^{2}$, Lixin Xu$^{3}$}

\affiliation{$^1$National Astronomy Observatories, Chinese Academy of Science, Beijing, 100012, People's Republic of China\\
$^2$Institute of Cosmology and Gravitation, University of Portsmouth, Portsmouth, PO1 3FX, United Kingdom\\
$^3$School of Physics and Optoelectronic Technology, Dalian University of Technology, Dalian, Liaoning 116024, People's Republic of China}

\begin{abstract}
We present improved constraints on an interacting vacuum model using \dw{updated} astronomical observations including the first data release from Planck. We consider a model with one dimensionless parameter, $\alpha$, describing the interaction between dark matter and vacuum energy (with fixed equation of state $w=-1$). The background dynamics correspond to a generalised Chaplygin gas cosmology, but the perturbations have a zero sound speed. The \dw{tension between the value of the Hubble constant, $H_0$, determined by Planck data plus WMAP polarisation (Planck+WP) and that determined by the Hubble Space Telescope (HST) can be alleviated by energy transfer from dark matter to vacuum ($\alpha>0$). A positive $\alpha$ increases the allowed values of $H_0$ due to parameter degeneracy within the model using only CMB data.} Combining with additional datasets of \dw{including} supernova type Ia (SN Ia) and baryon acoustic oscillation (BAO), we can significantly tighten the bounds on $\alpha$. Redshift-space distortions (RSD), which constrain the linear growth of structure, provide the tightest constraints on vacuum interaction when combined with Planck+WP, and prefer energy transfer from vacuum to dark matter ($\alpha<0$) which suppresses the growth of structure. Using the combined datasets of Planck+WP+Union2.1+BAO+RSD, we obtain the constraint on $\alpha$ to be $-0.083<\alpha<-0.006$ (95\% C.L.), allowing low $H_0$ consistent with the measurement from 6dF Galaxy survey. This interacting vacuum model can alleviate the tension between RSD and \dw{Planck+WP in the $\Lambda$CDM model for $\alpha<0$, or between HST measurements of $H_0$ and Planck+WP for $\alpha>0$, but not both at the same time.}
\end{abstract}


\maketitle

\section{Introduction}

One of the biggest challenges in modern cosmology is to explain the apparent accelerated expansion of the Universe today \cite{Riess:1998May, Perlmuter:1999Dec}. A variety of possible explanations have been put forward \cite{Li:2011Mar, Yoo:2012Dec, Sami:2013Sep} including \tc{allowing for the existence of} dark energy in Einstein gravity \tc{and modification of general relativity}. Vacuum energy is possibly the simplest model of dark energy, without any new dynamical degrees of freedom and with a vacuum equation of state (EoS), $\check{P}=-\check{\rho}=-V$. In Einstein gravity, a covariantly conserved vacuum energy density, $\nabla_\mu V=0$, is equivalent to a cosmological constant, $\Lambda=8\pi G_N V$. This is the basis of the $\Lambda$CDM cosmology, which is a highly predictive model to explain the present acceleration of the Universe. However, the $\Lambda$CDM model suffers from fine tuning and coincidence problems. As a result, many researchers have considered dynamical models of dark energy with a non-vacuum equation of state, $P\neq -\rho$, leading to a time-dependent dark energy density, e.g., scalar field \tc{models, \eg, quintessence \cite{quintessence}, phantom \cite{phantom}, quintom \cite{quintom}, or dark fluids} \dw{\cite{Kamenshchik:2001Mar, Bento:2002Feb,Piattella:2009kt}}. These different theories can be probed by a range of observational datasets \cite{Weinberg:2012Jan}.

In 2013 the Planck satellite provided a high-resolution measurement of \dw{anisotropies in the} \mv{cosmic microwave background} (CMB) \cite{Planck:2013Mar:map}. With the first release of Planck data, the cosmological \dw{analysis} from Planck collaboration showed that the standard spatially-flat $\Lambda$CDM model \dw{remains} an excellent fit to the \dw{CMB} data \cite{Planck:2013Mar:cos}. However, the results also pointed out some tension between Planck and other measurements of values of some cosmological parameters within the $\Lambda$CDM scenario \cite{Planck:2013Mar:cos}. Notably, the Planck collaboration presented a low value of the Hubble constant \footnote{\mvs{Besides the Planck data, there are also other observational estimations of the Hubble constant, which give a low value of $H_0$, see \eg, Ref.~\cite{Gott:2000Jun, Chen:2011May, Calabrese:2012May} using the median statistics method, Ref.~\cite{Beutler:2011Jun, Colless:2012Nov} from the 6dF Galaxy Survey, and Ref.~\cite{Busti:2014Feb, Verde:2014Mar} using Gaussian Processes by the measurements of $H(z)$.}}, $H_0=67.4\pm1.4$ km\,s$^{-1}$\,Mpc$^{-1}$ at 68\% C.L. from Planck data. When the sum of the masses of the active neutrinos is fixed to zero, the value of Hubble constant is changed slightly, giving $H_0=68.0\pm1.4$ km\,s$^{-1}$\,Mpc$^{-1}$. Both results from Planck data are \dw{in tension} with, for instance, direct measurements of $H_0$ by the Hubble Space Telescope (HST) observations of Cepheid variables, $H_0=73.8\pm2.4$ km\,s$^{-1}$\,Mpc$^{-1}$ \cite{Riess:2011Mar}. \dw{There is also some tension between the primary CMB anisotropies and measurements of the growth of structure, such as cluster number counts \cite{Ade:2013lmv}.}

This tension between the $H_0$ value determined from Planck and direct measurements of the Hubble constant \mvs{by HST} could be due to an incomplete understanding of the astrophysical observations. The direct measurements of $H_0$ have been revisited through reanalysing Cepheid data to address possible inconsistencies \cite{Efstathiou:2013Nov}. On the other hand, the determination on $H_0$ from CMB data is based on the assumption of an underlying theoretical model, so it is worthwhile to study the predictions in extensions of $\Lambda$CDM model, for instance, the neutrino $\Lambda$CDM model \cite{Wyman:2013Jul, Hamann:2013Aug, Battye:2013Aug}, dynamical dark energy models \cite{Xia:2013Aug}, or coupled dark energy models \cite{Salvatelli:2013Apr, Xia:2013Nov, Costa:2013Nov}.

A non-gravitational interaction between vacuum energy and matter provides an alternative framework in which to interpret the observational data. An interacting vacuum energy leads to a space- and time-dependent vacuum \cite{Wands:2012Mar, DeSantiago:2012Sep}, in which the gradient of the vacuum energy is given by a 4-vector,
\begin{eqnarray}
\nabla_{\mu}V=-Q_{\mu} \,.
\end{eqnarray}
The total energy-momentum must be conserved in a covariant theory, hence $Q_\mu$ describes the net energy-momentum transfer to the vacuum from other matter fields. \dw{Any} dark energy cosmology with exotic equation of state $P_X(\rho_X)$, can be decomposed \cite{Wands:2012Mar} into a cosmology with interacting vacuum energy density
\begin{equation}
\check\rho=-P_X \,,
\end{equation}
plus pressureless dark matter density
\begin{equation}
\rho_{\rm \tc{dm}}=\rho_X+P_X \,.
\end{equation}

In this paper we consider a cosmological model where the homogeneous background has the same behaviour as a generalised Chaplygin gas (GCG) \cite{Kamenshchik:2001Mar, Bento:2002Feb}. The GCG is parameterised by a single dimensionless parameter, $\alpha$, that in the interacting vacuum interpretation describes the energy transfer from matter to vacuum \cite{Bento:2004uh}. Thus we recover the $\Lambda$CDM model in the limit $\alpha\to0$. The original GCG model is severely constrained ($\alpha$ less than or of the order of $10^{-6}$) by large-scale structure formation since the barotropic dark fluid has a non-zero speed of sound for $\alpha\neq0$, which may lead to large oscillations, or instabilities, in the matter power spectrum \cite{Sandvik:2002Dec, Park:2009Oct}. Instead we will consider the interacting vacuum+matter model (a decomposed GCG model) where the energy-momentum transfer 4-vector is proportional to the matter 4-velocity. In this case there is no force on the dark matter particles in the dark matter rest frame and hence the dark matter follows geodesics. The sound speed of matter perturbations is zero and there are no oscillations in the matter power spectrum \cite{Wang:2013Jan, Borges:2013Sep}.

We revisit the constraints on this decomposed GCG using the new CMB data, including the temperature anisotropies from Planck \cite{Planck:2013Mar:cos} and polarization anisotropies from WMAP9 \cite{WMAP9:2012}. Firstly we focus on investigating the consistency between the CMB data alone and HST constraints on $H_0$. Then we perform the constraints on the interacting vacuum model using CMB data combined with other data. We use the updated baryon acoustic oscillations (BAO) data from the 6dF Galaxy Survey \cite{Beutler:2011Jun}, Sloan Digital Sky Survey (SDSS) DR7 \cite{Padmanabhan:2012Feb}, Baryon Oscillation Spectroscopic Survey (BOSS) DR9 \cite{Anderson:2012Mar}, and WiggleZ Dark Energy Survey \cite{Blake:2011Aug}. \tc{We also use the measurements of redshift space distortions (RSD) \cite{Percival:2004Jun, Blake:2011Apr, Samushia:2011Feb, Reid:2012Mar, Beutler:2012Apr}, which provides information of the growth of structure.}

This paper is organized as follows. In the next section, we review the interacting vacuum energy model and in particular the case of a decomposed GCG with geodesic flow. We examine the linear growth of structure and imprints on CMB power spectra in this model. In Section \ref{data}, we present the current observations and numerical analysis method. Then we show the results in Section \ref{results}. The conclusions and discussions are presented in Section \ref{summary}.

\section{Interacting vacuum energy model}

In a background cosmology with interacting vacuum energy, the Friedmann equation is given by
\begin{eqnarray}
 \label{Friedmann}
H^2=\frac{8 \pi G_N}{3}(\rho_b+\rho_r+\rho_{\rm \tc{dm}}+V) \,,
\end{eqnarray}
where $\rho_b$, $\rho_r$, $\rho_{\rm \tc{dm}}$ and $V$ are the energy densities of baryons, radiation, dark matter and interacting vacuum energy, and $H=\dot{a}/a$ is Hubble parameter.

For each component, the covariant conservation equation is written as
\begin{eqnarray}
 \label{covcon}
\nabla_\nu T^{\mu\nu}_{(I)}=Q^\mu_{(I)} \,,
\end{eqnarray}
where $Q^\mu_{(I)}=0$ for an independently-conserved component.
For interacting components, one conventionally splits the perturbed energy-momentum transfer 4-vector into the energy transfer, $Q_I+\delta Q_I$, and momentum transfer, $f^\mu_{(I)}$, relative to the total 4-velocity, $u^\mu$, \cite{Malik:2004Nov, Kodama:1984, Malik:2009Sep, Malik:2003Nov,Wands:2012Mar}
\begin{eqnarray}
\label{def:Qmu}
 Q^\mu_{(I)}=(Q_I+\delta Q_I)u^\mu+f^\mu_{(I)} \,.
\end{eqnarray}

At the background level, Eq.~(\ref{covcon}) reduces to the continuity equations for interacting vacuum and dark matter,
\begin{eqnarray}
 \label{rhodmdot}
\dot{\rho}_{\rm \tc{dm}}+3H\rho_{\rm \tc{dm}} &=&Q_{\rm \tc{dm}}= -Q \,, \\
 \label{Vdot}
\dot{V} &=& Q_{V}= Q \,.
\end{eqnarray}

\dw{We will work with the scalar perturbed Friedmann-Robertson-Walker (FRW) metric \cite{Malik:2009Sep}}
\begin{widetext}
\begin{eqnarray}
 ds^2 = -(1+2\phi)dt^2+2a\partial _i B dt dx^i+a^2 \left[(1-2\psi)\delta_{ij}+2\partial_i\partial_j E \right] dx^i dx^j \,.
\end{eqnarray}
\end{widetext}

In the linearly perturbed universe, \tc{the components of interacting vacuum and dark matter, Eq.~(\ref{covcon}) reduce to} the energy continuity equations
\begin{widetext}
\begin{eqnarray}
\label{dmdeltarho}
 \dot{\delta\rho}_{\rm \tc{dm}} +3H \delta\rho_{\rm \tc{dm}} -3 \rho_{\rm \tc{dm}} \dot\psi +\rho_{\rm \tc{dm}}\frac{\nabla^2}{a^2} \left(
 \theta_{\rm \tc{dm}} + a^2\dot{E} - aB \right)&=& -\delta Q - Q\phi \,,\\
 \label{Vdeltarho}
 \dot{\delta V} &=& \delta Q + Q\phi \,,
\end{eqnarray}
\end{widetext}
and the momentum conservation equations
\begin{eqnarray}
\label{dmtheta}
\rho_{\rm \tc{dm}}\dot\theta_{\rm \tc{dm}} + \rho_{\rm \tc{dm}}\phi &=& - f-Q(\theta-\theta_{\rm \tc{dm}}) \,,\\
\label{Vtheta}
-\delta V &=& f + Q\theta \,,
\end{eqnarray}
\mv{where $f=f_V=-f_{\rm dm}$.} Combining Eqs.~(\ref{dmdeltarho}) and (\ref{Vdeltarho}), Eqs.~(\ref{dmtheta}) and (\ref{Vtheta}) \dw{to eliminate $\delta Q$ and $f$}, we obtain \cite{Wands:2012Mar}
\begin{widetext}
\begin{eqnarray}
\label{finaldeltarho}
 \dot{\delta\rho}_{\rm \tc{dm}} + 3H \delta\rho_{\rm \tc{dm}} -3 \rho_{\rm \tc{dm}} \dot\psi + \rho_{\rm \tc{dm}}\frac{\nabla^2}{a^2} \left( \theta_{\rm \tc{dm}} + a^2\dot{E} - aB \right)
 &=& -\dot{\delta V} \,,\\
\label{finaltheta}
\rho_{\rm \tc{dm}}\dot\theta_{\rm \tc{dm}} + \rho_{\rm \tc{dm}}\phi &=& \delta V + Q \theta_{\rm \tc{dm}} \,.
\end{eqnarray}
\end{widetext}

\mv{\subsection{The decomposed GCG with geodesic flow}}

We apply the interacting vacuum energy to the GCG model with a unified EoS \cite{Kamenshchik:2001Mar, Bento:2002Feb},
\begin{eqnarray}
 \label{PgCg}
P_{\rm \tc{gCg}}=-\frac{A}{\rho_{\rm \tc{gCg}}^\alpha} \,,
\end{eqnarray}
\tc{where} $A$ and $\alpha$ are the unified GCG model parameters. We have,
\begin{eqnarray}
 \label{decompose}
\rho_{\rm \tc{gCg}} &=& \rho_{\rm \tc{dm}}+V,~~~~P_{\rm \tc{gCg}}=-V\,,
\end{eqnarray}
such that, from Eq.~(\ref{PgCg}),
\begin{eqnarray}
 \label{A}
  A&=&V(\rho_{\rm \tc{dm}}+V)^{\alpha} \,.
\end{eqnarray}
Note that the decomposed GCG model is characterised by an interaction parameter $\alpha$. Combining Eqs.~(\ref{rhodmdot}, \ref{Vdot}, \ref{decompose}, \ref{A}), we obtain \cite{Wands:2012Mar}
\begin{equation}
 \label{backQ}
 Q \dw{=\dot{V}} = 3\alpha H\frac{\rho_{\rm \tc{dm}} V}{\rho_{\rm \tc{dm}} +V} \,.
\end{equation}

\begin{figure}[!htbp]
\includegraphics[scale=0.3]{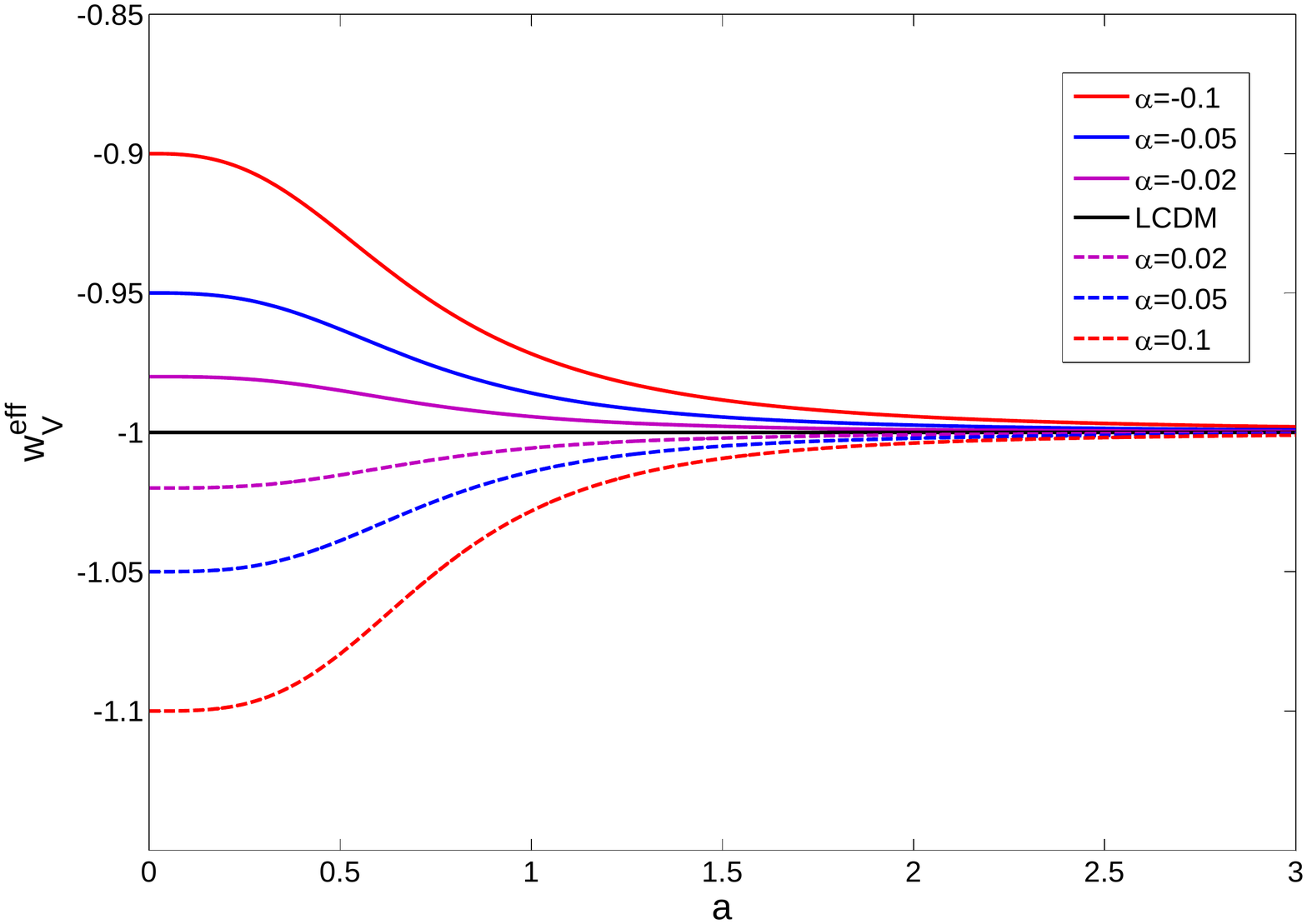} \\
  \caption{The time evolution of the effective EoS of dark energy for different $\alpha$ values.}\label{fig:weff}
\end{figure}

With the above expression, rewriting Eqs.~(\ref{rhodmdot}, \ref{Vdot}) as
\begin{eqnarray}
  \dot{\rho}_{\rm \tc{dm}} +3H(1+w_{\rm \tc{dm}}^{\rm \tc{eff}})\rho_{\rm \tc{dm}} &=& 0 \,, \\
  \dot{V}+3H(1+w_V^{\rm \tc{eff}})V &=& 0 \,,
\end{eqnarray}
we \tc{can derive the effective EoS for dark matter and vacuum respectively as,}
\begin{eqnarray}
  w_{\rm \tc{dm}}^{\rm \tc{eff}}&=&\alpha\frac{V}{\rho_{\rm \tc{dm}}+V} \,, \\
  \label{weff}
  w_V^{\rm \tc{eff}}&=&-1-\alpha\frac{\rho_{\rm \tc{dm}}}{\rho_{\rm \tc{dm}}+V} \,.
\end{eqnarray}
We find that a non-zero interaction indicates the effective dark matter component with $w_{\rm \tc{dm}}^{\rm \tc{eff}}\neq0$, while the effective dark energy behaves like a quintessence for \tc{a} negative $\alpha$ or a phantom for \tc{a} positive $\alpha$. A constant interaction parameter, $\alpha$, cannot realize a quintom-like effective dark energy with the EoS crossing $-1$. In the limits at early times, when $\rho_{\rm \tc{dm}}\gg V$, we have $w_{\rm \tc{dm}}^{\rm \tc{eff}}\rightarrow 0$ and $w_V^{\rm \tc{eff}}\rightarrow -1-\alpha$, where the model becomes the $w$CDM model with a constant EoS. \tc{In the future limit where dark matter is diluted away, }we have $w_{\rm \tc{dm}}^{\rm \tc{eff}}\rightarrow \alpha$ and $w_V^{\rm \tc{eff}}\rightarrow -1$. The evolution of \tc{the} effective EoS of dark energy is shown in Fig.~\ref{fig:weff}.

At the linear perturbation level, we consider an energy flow parallel to the 4-velocity of dark matter
\begin{equation}
 \label{Qmu}
 Q^{\mu}_{(\rm \tc{dm})} = -Qu^{\mu}_{(\rm \tc{dm})} \,.
\end{equation}
\dw{In this case the dark matter} follows geodesics \cite{Wands:2012Mar,DeSantiago:2012Sep,Wang:2013Jan}. It means that the vacuum energy perturbations vanish in the dark matter-comoving frame, from Eq.~(\ref{finaltheta})
\dw{
\begin{equation}
\delta V+\dot{V}\theta_{\rm \tc{dm}} =0 \,.
\end{equation}
}In this case, the spatial hypersurface orthogonal to the dark matter 4-velocity coincides with that orthogonal to the vacuum energy flow \cite{Wands:2012Mar, DeSantiago:2012Sep}. It is noted that for dark energy with constant EoS $w\neq-1$, with the same covariant interaction, Eq.~(\ref{Qmu}), there are inhomogeneous dark energy \cite{Li:2013Dec}.

\dw{We will work in a synchronous gauge ($\phi=B=0$) where $h$ characterizes a scalar mode of spatial metric perturbations. The dark matter momentum conservation (\ref{finaltheta}) then requires
\begin{equation}
\rho_{\rm dm} \dot\theta_{\rm dm} =0 \,.
\end{equation}
}
In order to fix the residual gauge freedom in \tc{the} synchronous gauge \cite{Ma:1995Jun} we take $\theta_{\rm \tc{dm}}=0$, \dw{and thus $\delta V=0$}. In \tc{the} comoving synchronous gauge, the density perturbation equation (\ref{finaldeltarho}) for dark matter then has the simple form
\begin{eqnarray}
 \label{dmenergy}
 \dot{\delta}_{\rm \tc{dm}}=-\frac{\dot{h}}{2}+\frac{Q}{\rho_{\rm \tc{dm}}}\delta_{\rm \tc{dm}} \,.
\end{eqnarray}

For the non-interacting baryon component, the perturbation equations for the baryon density contrast and velocity after decoupling are given by,
\begin{eqnarray}
 \label{benergy-mom}
 \dot{\delta}_b =\frac{k^2}{a^2}\theta_b-\frac{\dot{h}}{2} \,,~~~\dot\theta_b = 0 \,.
\end{eqnarray}

\subsection{Linear growth of structure}

From Eq.~(\ref{dmenergy}), we can see that the interaction has a direct effect on the dark matter density perturbations. The detailed discussions of the effects of the interaction parameter on CMB and large-scale structure power spectra \tc{are} given in Ref.~\cite{Wang:2013Jan}. In the following, we \tc{shall} investigate the linear growth rate of structure formation in the interacting vacuum model.

Generally, for the $\Lambda$CDM model in Einstein gravity the matter overdensity obeys the ordinary second-order differential equation
\begin{eqnarray}
\label{mtwo}
 \ddot{\delta}_{\rm \tc{m}}+2H\dot{\delta}_{\rm \tc{m}} = 4 \pi G_N \rho_{\rm \tc{m}} \delta_{\rm \tc{m}} \,.
\end{eqnarray}

For the interacting vacuum model, combining Eqs.~(\ref{dmenergy}, \ref{benergy-mom}) with field equation in the synchronous gauge:
\begin{eqnarray}
\label{htwodot}
 \ddot{h}+2H\dot{h} = -8 \pi G_N (\delta\rho+3\delta P) \,,
\end{eqnarray}
we can obtain the second-order differential equations for dark matter overdensity and baryon overdensity \tc{respectively as,}
\begin{widetext}
\begin{eqnarray}
\label{dmtwo}
 \ddot{\delta}_{\rm \tc{dm}}+\left(2H-\frac{Q}{\rho_{\rm \tc{dm}}}\right)\dot{\delta}_{\rm \tc{dm}}-\left[2H\frac{Q}{\rho_{\rm \tc{dm}}}+\dot{\left(\frac{Q}{\rho_{\rm \tc{dm}}}\right)}\right]
 \delta_{\rm \tc{dm}}=4\pi G_N (\rho_{\rm \tc{dm}}\delta_{\rm \tc{dm}}+\rho_{b}\delta_{b}) \,, \\
 \label{baryontwo}
 \ddot{\delta}_b+2H\dot{\delta}_b = 4 \pi G_N (\rho_{\rm \tc{dm}}\delta_{\rm \tc{dm}}+\rho_{b}\delta_{b}) \,.
\end{eqnarray}
\end{widetext}
The total components on the right-hand side of Eq.~(\ref{htwodot}) are the sum of dark matter and baryons when matter domination starts. Note that \dw{in a comoving-synchronous gauge we have a spatially homogeneous vacuum energy, $\delta V=0$.}

Defining function, $g_I(a)\equiv\delta_I(a)/a$, and replacing the variable $t$ by $x=\ln a$, we can obtain
\begin{widetext}
\begin{eqnarray}
\label{dmtwox}
 g_{\rm \tc{dm}}''+\left[3+(\ln \mathcal{H})'-\frac{aQ}{\mathcal{H}}\right]g_{\rm \tc{dm}}'+\left[2+(\ln \mathcal{H})'-3\frac{aQ}{\mathcal{H}}-\frac{aQ'}{\mathcal{H}}\right]g_{\rm \tc{dm}}= \frac{4 \pi G_N a^2}{\mathcal{H}^2}(\rho_{\rm \tc{dm}}g_{\rm \tc{dm}}+\rho_{b}g_{b})  \,, \\
 \label{baryontwox}
 g_b''+\left[3+(\ln \mathcal{H})'\right]g_b'+\left[2+(\ln \mathcal{H})'\right]g_b= \frac{4 \pi G_N a^2}{\mathcal{H}^2}(\rho_{\rm \tc{dm}}g_{\rm \tc{dm}}+\rho_{b}g_{b})  \,,
\end{eqnarray}
\end{widetext}
where the primes denote the derivations with respect to $x$, and $\mathcal{H} = aH$ is the conformal Hubble parameter. Correspondingly, the \dw{overall} growth rate of matter is
\begin{equation}
\label{fm}
f_{\rm \tc{m}}(a) \equiv [\ln \delta_{\rm \tc{m}}(a)]'=1+g_{\rm \tc{m}}'(a) \,,
\end{equation}
\dw{where}
\begin{equation}
\label{gm}
g_{\rm \tc{m}}(a) = \frac{\rho_{\rm \tc{dm}}}{\rho_{\rm \tc{m}}}g_{\rm \tc{dm}}+\frac{\rho_b}{\rho_{\rm \tc{m}}}g_b \,.
\end{equation}
Then we can obtain the \tc{evolution of $f_{\rm \tc{m}}$ with redshift $z$} by numerically solving the closed \tc{differential equation set, Eqs.~}(\ref{dmtwox}, \ref{baryontwox}), with the initial \tc{\mv{condition set} in the matter-domination era}, $g_I(a_i)=1$ and $g'_I(a_i)=0$. In a particular case \tc{in which} $\alpha=0$, the solutions of Eqs.~(\ref{dmtwox})-(\ref{gm}) are identical to that of the original Eq.~(\ref{mtwo})\tc{, which means that the interacting vacuum model reduces to the $\Lambda$CDM model when $\alpha=0$ at the linear perturbation level.}

\begin{figure}[!htbp]
\includegraphics[scale=0.3]{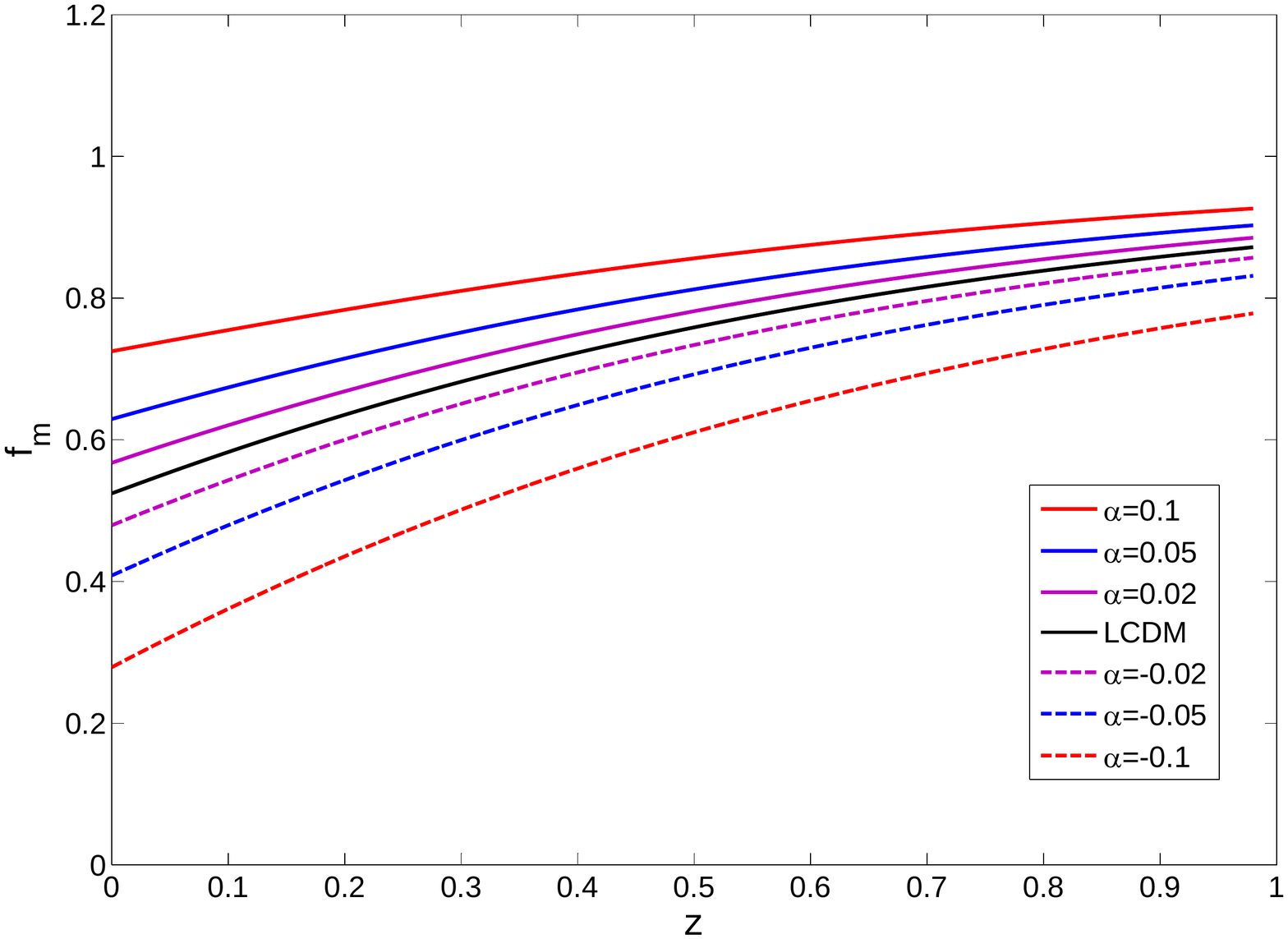} \\
  \caption{The time evolution of $f_{\rm \tc{m}}$ for different $\alpha$ values, \dw{with the same value of $\Omega_m$ today}.}\label{fig:fz}
\end{figure}

The evolution of the growth rate, $f_{\rm \tc{m}}$, is shown in Fig.~\ref{fig:fz}, from which we can see that a positive interaction leads to a faster growth than that in a $\Lambda$CDM model \dw{with the same $\Omega_{m}$ today.}. This is due to \tc{the} energy transfer from dark matter to vacuum energy for \tc{a} positive $\alpha$. Conversely, we obtain a slower growth for a negative $\alpha$ than that in $\Lambda$CDM.

\tc{Observationally, the quantity of $f\sigma_8$, can be measurement from redshift surveys using the RSD effect \cite{Song:2008Jul, Percival:2008Aug, Samushia:2012Jun, Xu:2013Feb}.} Here $\sigma_8(z)$ is the root mean square (rms) amplitude of density fluctuations in a sphere of comoving radius $R_8=8$ h$^{-1}$\,Mpc,
\begin{eqnarray}
\sigma_8(z)=\left\{\frac{1}{2\pi^2}\int_0^\infty dk k^2 W^2_8(k)P(k, z)\right\}^{1/2}\,,
\end{eqnarray}
where $W_8(k)$ is the Fourier transform of the top-hat window function with the width of the comoving scale $R_8$ and $P(k, z)$ is the matter power spectrum.

\begin{figure}[!htbp]
\includegraphics[scale=0.3]{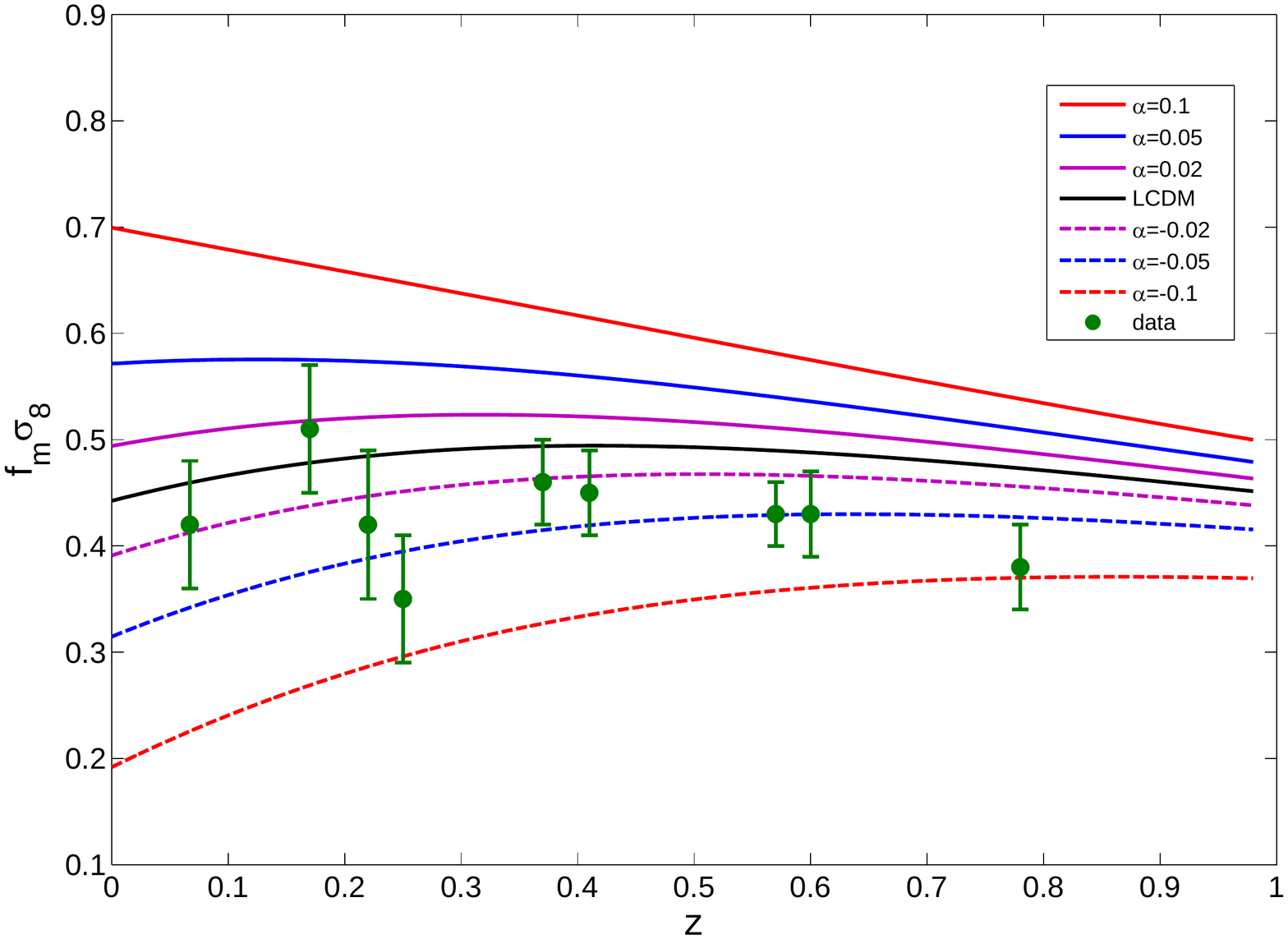} \\
\includegraphics[scale=0.3]{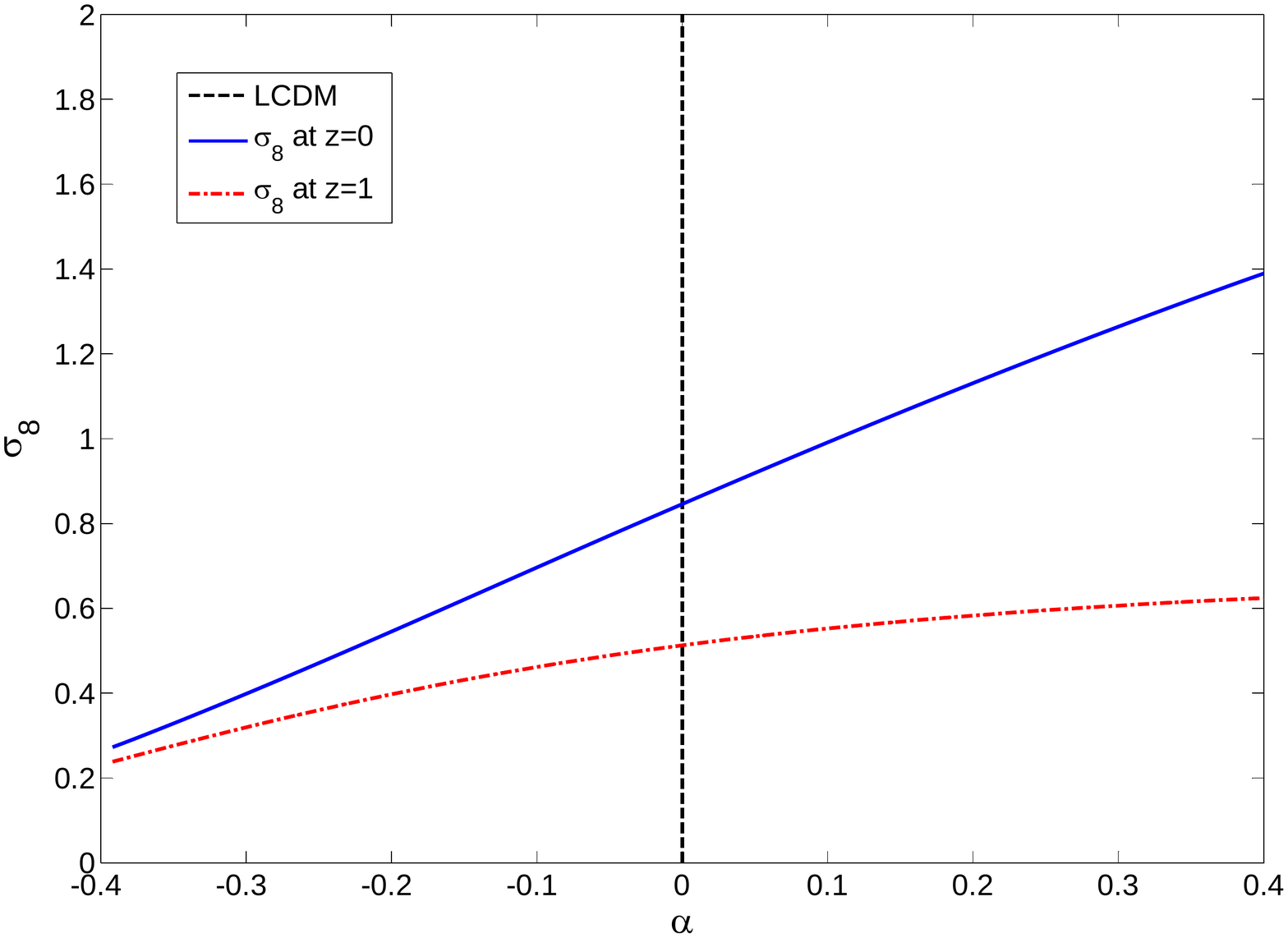} \\
  \caption{\tc{Upper panel: time evolution of $f_{\rm \tc{m}}(z)\sigma_8(z)$ for different values of $\alpha$, \dw{with the same value of $\Omega_m$ today}. The points with error bars are the observational data \cite{Percival:2004Jun, Blake:2011Apr, Samushia:2011Feb, Reid:2012Mar, Beutler:2012Apr}, summarized in Ref.~\cite{Samushia:2012Jun}; Lower panel: predictions for $\sigma_8$ with the interaction parameter, $\alpha$, at the given redshifts.}}\label{fig:f8-sigma8}
\end{figure}

In order to test the interacting vacuum model \tc{using the} RSD data, we firstly illustrate the theoretical predictions of $f_{\rm \tc{m}}(z)\sigma_8(z)$ for different interaction parameters, as shown in the \tc{upper} panel of Fig.~\ref{fig:f8-sigma8}. It is found that there \tc{exist obvious dispersions} of $f_{\rm \tc{m}}(z)\sigma_8(z)$ due to different values of interaction parameter, $\alpha$. For $\alpha > 0$, $f_{\rm \tc{m}}(z)\sigma_8(z)$ is larger than that in the $\Lambda$CDM model. As we mentioned above, the growth rate for a positive interaction becomes larger than that in the $\Lambda$CDM model, because of an energy transfer from dark matter to vacuum energy. A positive interaction shifts the matter-radiation equality to an earlier time, which yields a higher value of $\sigma_8$. The plot in the \tc{lower} panel of Fig.~\ref{fig:f8-sigma8} shows that at a given redshift the larger value of $\alpha$ is, the higher value of $\sigma_8$ is. Conversely, $f_{\rm \tc{m}}(z)\sigma_8(z)$ is suppressed for \tc{a} negative $\alpha$, compared with that in the $\Lambda$CDM model.

\tc{Moreover, $\alpha$ has a larger impact on $f_{\rm \tc{m}}(z)\sigma_8(z)$ at a later time, \ie, models with different $\alpha$'s differ the most at $z=0$. This is because both $f_{\rm \tc{m}}$ and $\sigma_8$ are affected more as time evolves, as illustrated in Figs.~\ref{fig:fz} and \ref{fig:f8-sigma8} (bottom panel) respectively.}

\subsection{Imprints on CMB}

\begin{figure}[!htbp]
\includegraphics[scale=0.3]{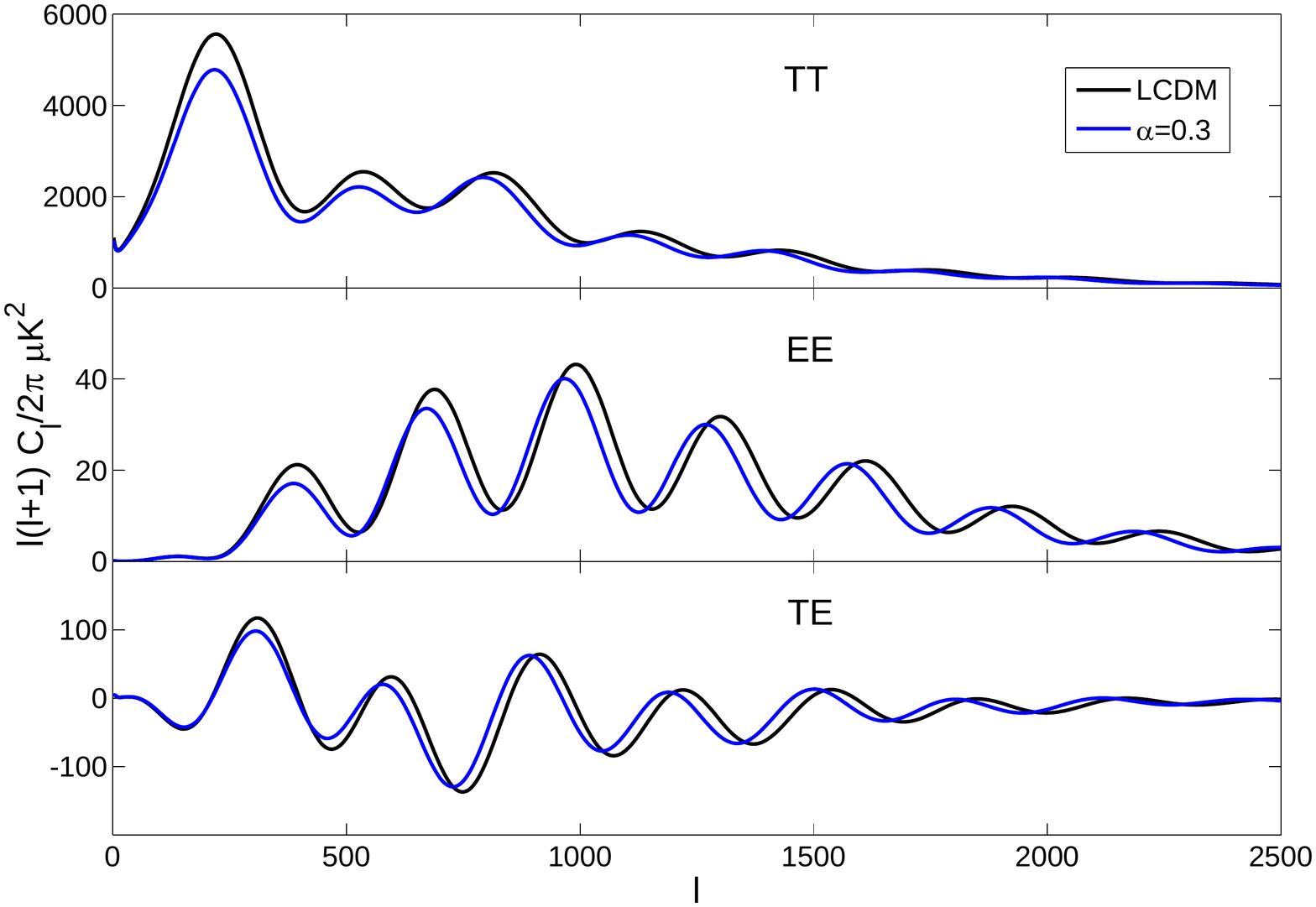} \\
\includegraphics[scale=0.3]{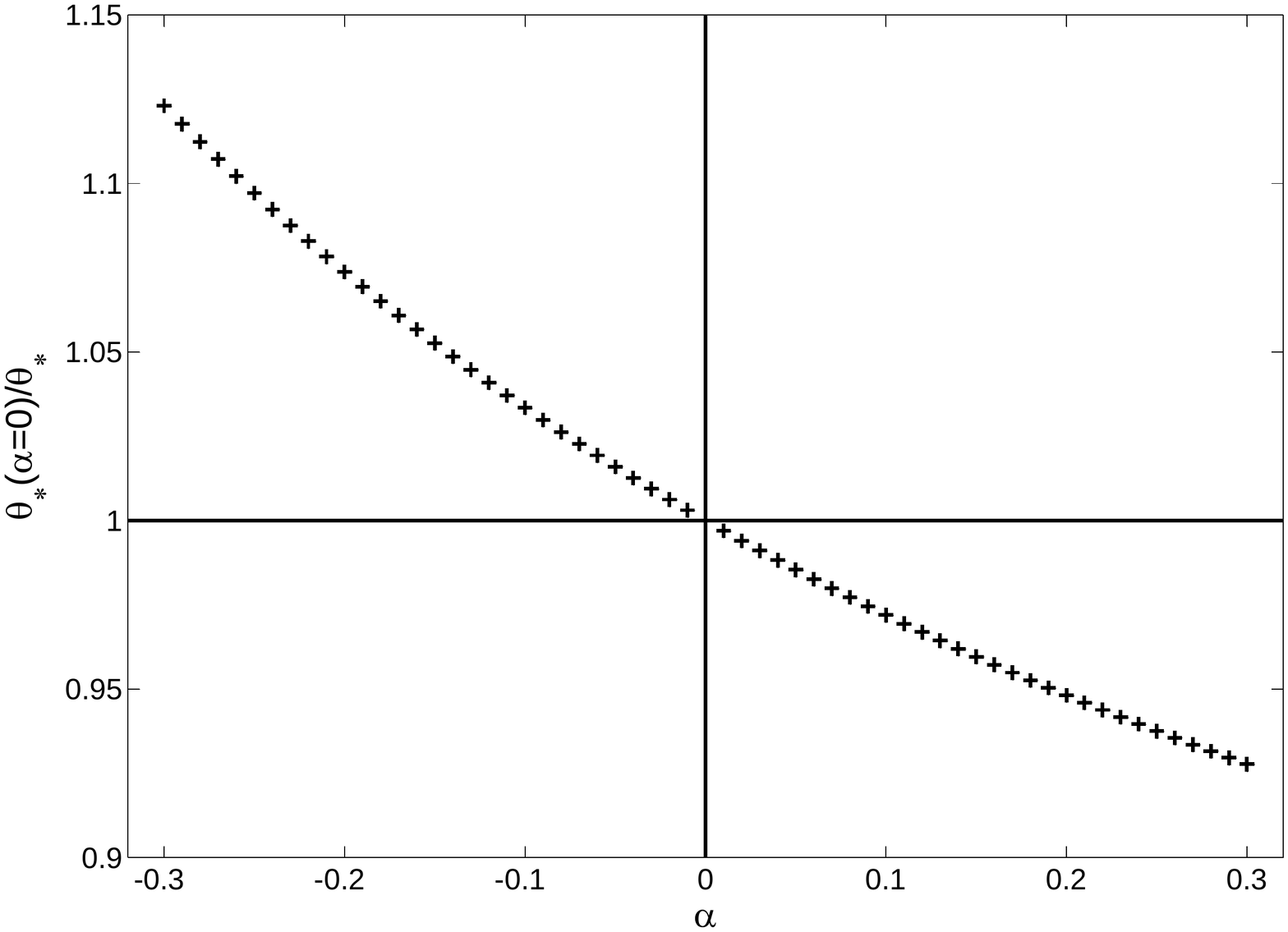} \\
  \caption{Upper panel: CMB TT(top), EE(middle) and TE(bottom) power spectra \tc{for $\alpha=0.3$ (blue) and $\Lambda$CDM (black) models}; Lower panel: the ratio of $\theta_{\ast}$ between the $\Lambda$CDM model and interacting vacuum model as a function of $\alpha$.}\label{fig:Cl}
\end{figure}

\tc{We take the $\alpha=0.3$ model for an example to show the imprint of the interacting vacuum energy model on CMB power spectra in the upper panel of Fig.~\ref{fig:Cl}. \dw{Positive $\alpha$ suppresses the height of the peaks as it increases the matter density at the time of recombination. $\alpha>0$} also shifts the location of the peaks to the low-$\ell$ end. To understand the physics, in the lower panel of Fig.~\ref{fig:Cl} we show the ratio of $\theta_{\ast}$ (the ratio of the sound horizon to angular diameter distance at last-scattering) for the $\Lambda$CDM model over the interacting vacuum model, and we can see that this ratio decreases monotonically with $\alpha$. At $\alpha=0.3$, $\theta_{\ast}$ is about 7\% larger than that of the $\Lambda$CDM model. Note that the $n$th CMB acoustic peak is \dw{approximately located} at}
\tc{
\begin{equation}
\ell=\frac{n\pi}{\theta_{\ast}}
\end{equation}
}\tc{So the CMB peaks of the $\alpha=0.3$ model \dw{appear} at slightly smaller $\ell$'s compared to the $\Lambda$CDM model, and this is what we have seen in the CMB power spectra. }

\section{Data and Method} \label{data}

In this section, we \tc{describe the data sets we use and analysis methods we adopt.}

Here are the current observations we used:

\begin{enumerate}
\item The recently released Planck data include the high-\tc{$\ell$} temperature power spectrum from the CAMSpec likelihood with a wide multipole range covering from $\tc{\ell}=50$ to $\tc{\ell}=2500$, and the low-\tc{$\ell$} temperature power spectrum from the Commander likelihood over the multipole range $\tc{\ell}=2-49$ \cite{Planck:2013Mar:cos}.
\dw{We set the Planck lensing parameter $A_L=1$; that is, we use the full information from the high-$\ell$ power spectrum including the effect of gravitational lensing along the line of sight on the temperature anisotropies.}
\tc{So far the Planck team has only provided} the temperature power spectrum data. The low-\tc{$\ell$} polarization power spectra (up to $\tc{\ell}=32$) are from WMAP 9-year data \cite{WMAP9:2012}. So the combination of CMB temperature data from Planck and polarization data from WMAP9 is denoted as ``Planck+WP". When performing constraints \tc{using} Planck data, extra $14$ foreground parameters are allowed to vary freely. \tc{For comparison, we do the same analysis using both CMB temperature and polarization anisotropies from WMAP 9-year data, which is denoted as ``WMAP9".}

\item  A Gaussian likelihood function of $H_0=73.8\pm2.4$ km\,s$^{-1}$\,Mpc$^{-1}$ from HST \cite{Riess:2011Mar} is taken when the direct measurement data is included. \mvs{To make a comparison between the effects of different $H_0$ priors, we also use another Gaussian prior on the Hubble constant with a relatively low value, $H_0=67\pm3.2$ km\,s$^{-1}$\,Mpc$^{-1}$ from 6dF Galaxy survey \cite{Beutler:2011Jun, Colless:2012Nov}, labeled ``lowH".}

\item  For SN Ia data, we use Union2.1 compilation of 580 SN Ia with systematic errors \cite{Suzuki:2011May}. \footnote{\tc{Note that another supernova sample, referred to SNLS compilation \cite{Conley:2011Apr}, has been recalibrated with an improved accuracy using the cross-calibration between SNLS and SDSS supernova samples \cite{Betoule:2012Dec}. In \tc{a} recent work \cite{Betoule:2014Jan}, the updated cosmological constraints has been presented using the combined SNLS 3-year data and the full SDSS-II spectroscopic sample from the final release of the SDSS-II supernova survey \cite{Sako:2014Jan}. The cosmological parameters derived from this sample is similar to that from the Union2.1 sample, which is used in this work.}}

\item  BAO measurement in the matter power spectrum is regarded as a cosmic standard ruler and helps strengthen observational constraints on cosmological parameters. Usually, an effective distance quantity, $D_V(z)$, is used, which contains both the angular-diameter distance, $D_A(z)$, and the expansion history, $H(z)$. $D_V(z)$ is expressed as
\begin{eqnarray}
D_V(z)=[(1+z)^2D^2_A(z)cz/H(z)]^{1/3} \,.
\end{eqnarray}
The updated BAO data include $D_V(0.106)=456\pm27 $ Mpc from 6dF Galaxy Survey \cite{Beutler:2011Jun}; the distance ratio of the effective distance, $D_V(z)$, to the comoving sound horizon at the baryon drag epoch, $r_s(z_{\rm \tc{drag}})$, at $D_V(0.35)/r_s=8.88\pm0.17$ from SDSS DR7 data \cite{Padmanabhan:2012Feb}, denoted as ``SDSS(R)" and $D_V(0.57)/r_s=13.62\pm0.22$ from BOSS DR9 data \cite{Anderson:2012Mar}, denoted as ``BOSS". In addition, WiggleZ Dark Energy survey obtain BAO feature by the acoustic parameter, $A(z)$, which is related to $D_V(z)$ by
\begin{eqnarray}
A(z)=D_V(z)\sqrt{\Omega_{\rm \tc{m}} H_0^2}/cz \,,
\end{eqnarray}
at $A(0.44)=0.474\pm0.034, A(0.60)=0.442\pm0.020$, and $A(0.73)=0.424\pm0.021$ \cite{Blake:2011Aug}.

\item RSD is one of the main sources of anisotropy in galaxy power spectra caused by the peculiar velocities of galaxies. The observations of RSD in terms of $f_{\rm \tc{m}}(z)\sigma_8(z)$ provide a good way to measure the \dw{linear growth of structure in the Universe}. We use 9 data points \cite{Percival:2004Jun, Blake:2011Apr, Samushia:2011Feb, Reid:2012Mar, Beutler:2012Apr}, compiled in the Table 1 of Ref.~\cite{Samushia:2012Jun}. Here an old data point at the redshift $z=0.77$ from the VVDS is replaced by \tc{a recent measurement }at a very similar redshift $z=0.78$ from WiggleZ \cite{Samushia:2012Jun}.

\end{enumerate}

We test the interacting vacuum model against these observations using \tc{a modified version of the} CosmoMC packge \cite{ref:MCMC, Lewis2013Apr}.
The set of cosmological parameters allowed to vary and their corresponding top-hat priors we adopted are: the physical baryon density, $\Omega_{b}h^2 \in [0.005, 1]$, the physical dark matter density, $\Omega_{\rm \tc{dm}}h^2 \in [0.001, 0.99]$,  \tc{the ratio ($\times$ 100) of the sound horizon to angular diameter distance at last-scattering}, $\Theta_S \in [0.5, 1.5]$, the optical depth, $\tau \in [0.01, 0.2]$, the scalar spectral index of the primordial power spectrum, $n_s \in [0.5, 1.5]$ and the amplitude of the primordial power spectrum $\log[10^{10}A_s] \in [2.7, 4]$ with the pivot scale, $k_{s0} = 0.05$ ${\rm Mpc}^{-1}$.

In the interacting vacuum model with geodesic flow, we have a zero rest-frame sound speed. The interaction parameter \tc{is allowed to vary from negative} to positive values. A negative $\alpha$ leads to energy transfer from vacuum energy to dark matter. In order to guarantee that the evolution of Universe undergoes the domination transition from matter to dark energy, we need to consider possible limits on negative ranges of $\alpha$. Based on the ratio of dark matter density to vacuum energy density,
\begin{equation}\label{ratio}
  \frac{\rho_{\rm \tc{dm}}}{V} \propto a^{-3(1+\alpha)},
\end{equation}
we can find that for $\alpha > -1$, the ratio decreases with \tc{time}. Conversely, the $\alpha$'s values less than $-1$ make the ratio become larger and larger with \tc{time}, which is obviously inconsistent with current observations. Therefore, we set a \tc{flat} prior for the interaction parameter, \tc{namely}, $\alpha \in (-0.99, 1.5]$. In addition, we fix the sum of the masses of three active neutrinos to $\sum m_{\nu} = 0$ and the effective number of neutrino species to $N_{\rm \tc{eff}} = 3.046$.

The convergence of Markov chains is tested by the Gelman and Rubin criterion. Here the $R-1$ value is required to be below $0.03$.

\section{Results} \label{results}

\begin{table*}
\begin{center}
\begin{tabular}{|c|c|c|c|c|c|c|c|c|c|}
\hline \hline
Models&\multicolumn{4}{|c|}{$\Lambda$CDM}&\multicolumn{5}{|c|}{Interacting vacuum}\\
\hline
Parameters & $H_0$ & $\Omega_{\rm \tc{m}}$ &$\Omega_{\rm \tc{m}} h^3$ &$\sigma_8$ &$~~~~~~~~\alpha~~~~~~~
$& $H_0$ & $\Omega_{\rm \tc{m}}$ &$\Omega_{\rm \tc{m}} h^3$ & $\sigma_8$\\ \hline
 WMAP9      &  $70.3_{-        2.2}^{+        2.2}$ &
               $0.277_{-        0.028}^{+        0.023}$ &
               $0.0957_{-        0.0017}^{+        0.0018}$&
               $0.821_{-        0.023}^{+        0.023}$ &
               $~~0.119_{-        0.343}^{+        0.260}$ &
               $72.1_{-        6.2}^{+        7.2}$&
               $0.254_{-        0.166}^{+        0.076}$&
               $0.0856_{-        0.0162}^{+        0.0287}$&
               $1.013_{-        0.441}^{+        0.154}$\\ \hline
 Planck+WP &   $ 68.0_{-        1.2}^{+        1.2}$&
               $0.307_{-        0.018}^{+        0.016}$ &
               $0.0962_{-        0.0006}^{+        0.0006}$&
               $0.840_{-        0.013}^{+        0.013}$ &
               $-0.021_{-        0.294}^{+        0.215}$&
               $67.0_{-        5.5}^{+        5.5}$&
               $0.342_{-        0.172}^{+        0.101}$ &
               $0.0959_{-        0.0091}^{+        0.0218}$&
               $0.868_{-        0.284}^{+        0.095}$ \\
\hline\hline
\end{tabular}
\caption{Comparison of some parameters at 68\% C.L. for the $\Lambda$CDM model and interacting vacuum model from the constraints of only CMB data: WMAP9 and Planck+WP.}\label{tab:9-pl-only-results}
\end{center}
\end{table*}

\begin{figure}[!htbp]
\includegraphics[scale=0.4]{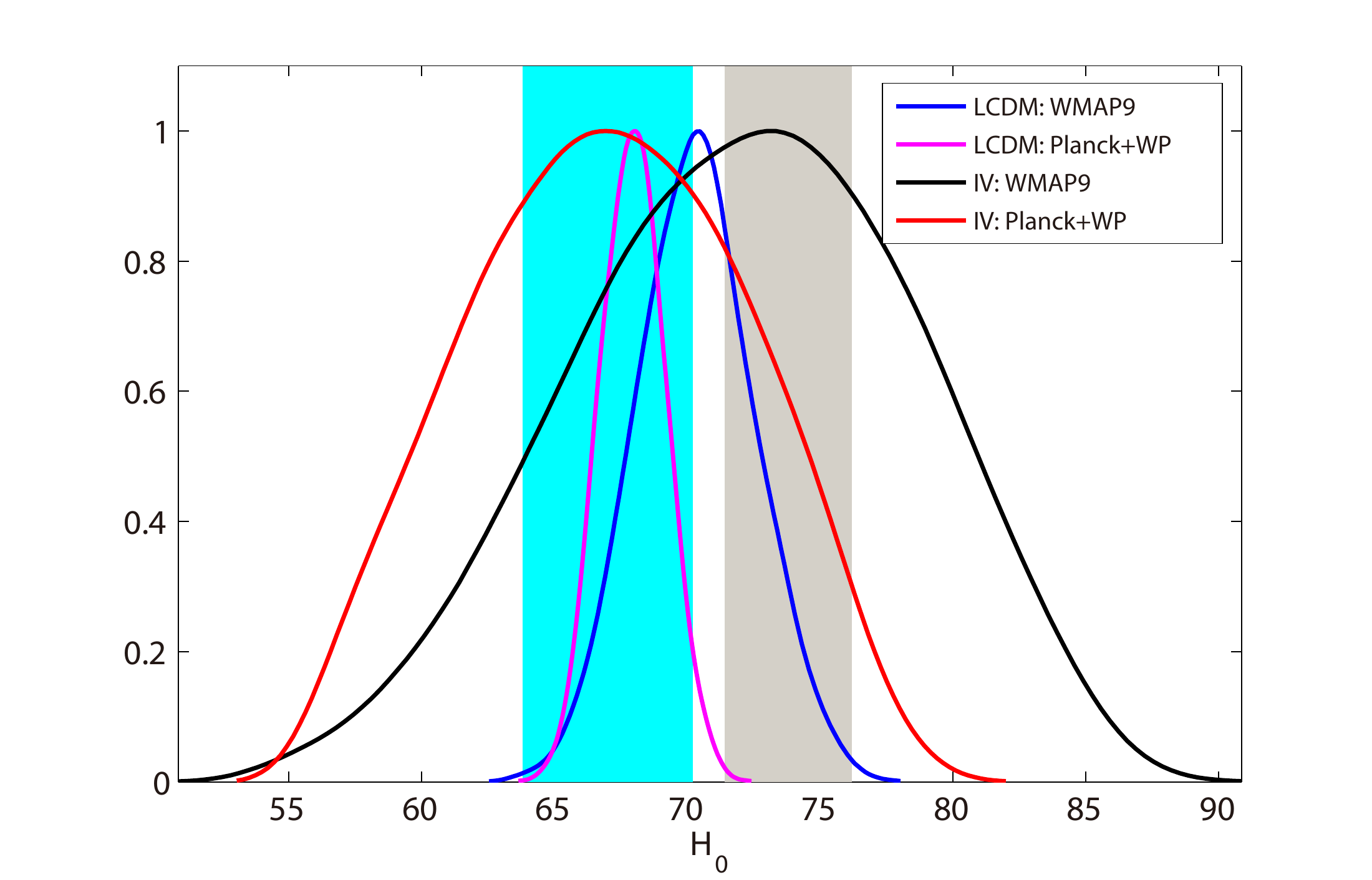}  \\
  \caption{The 1D marginalized distributions of $H_0$ for the $\Lambda$CDM model and interacting vacuum (represented by ``IV" for short in the plot) model from the constraints of only CMB data: WMAP9 and Planck+WP. The grey band corresponds to the direct measurement of $H_0$ from HST at the 68\% C.L. \mvs{The cyan band is the $1\,\sigma$ range of $H_0$ measured by 6dF Galaxy Survey}.}\label{fig:1D-H0-com}
\end{figure}

\begin{figure*}[!htbp]
\includegraphics[scale=0.45]{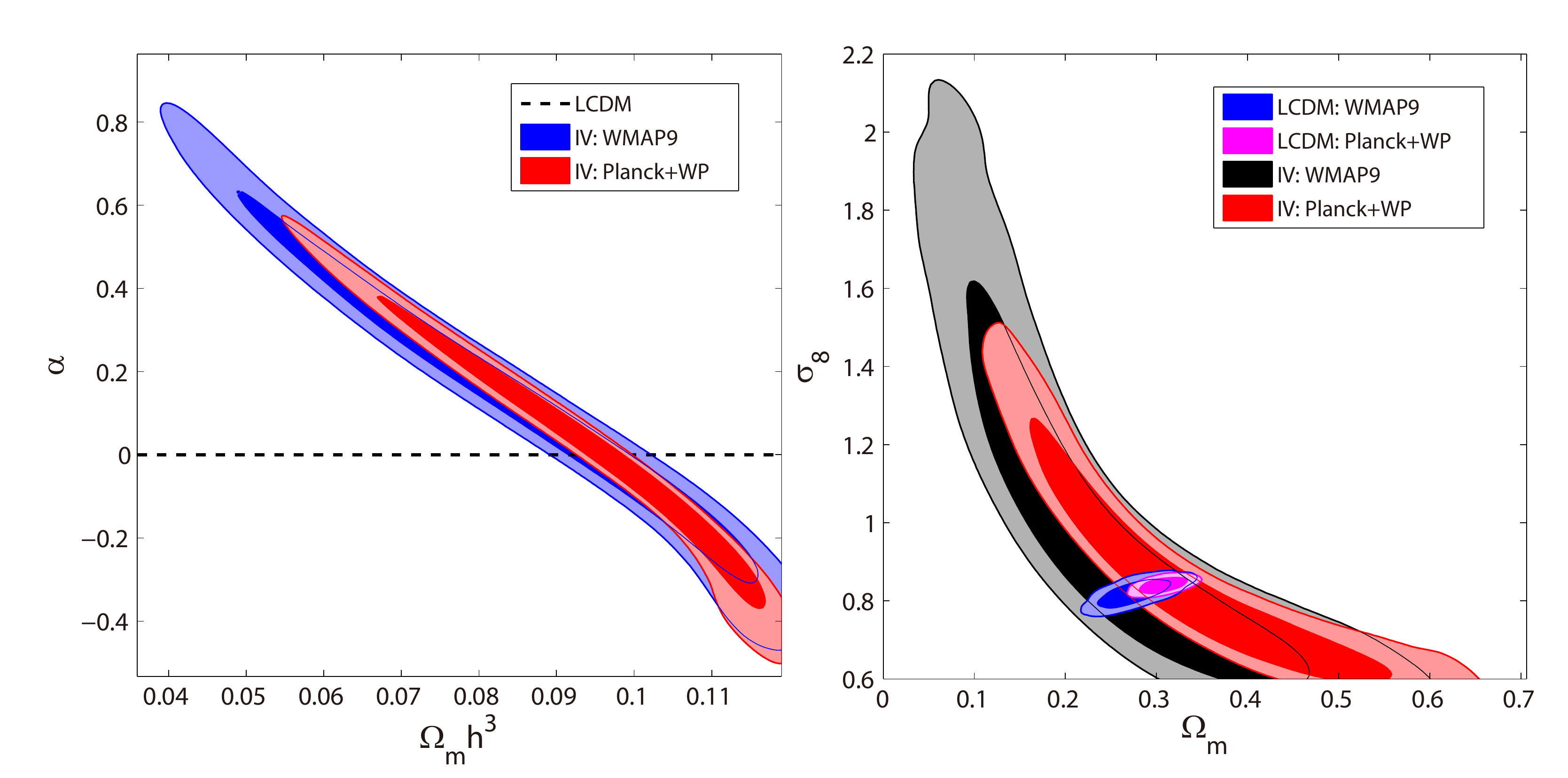} \\
  \caption{Left panel: the 2D contours of $\Omega_{\rm \tc{m}}h^3-\alpha$ at the 68\% C.L. and 95\% C.L. Right panel: the 2D contours of $\Omega_{\rm \tc{m}}-\sigma_8$ at the 68\% C.L. and 95\% C.L.}\label{fig:alpha-omegah3}
\end{figure*}

\begin{figure*}[!htbp]
\includegraphics[scale=0.45]{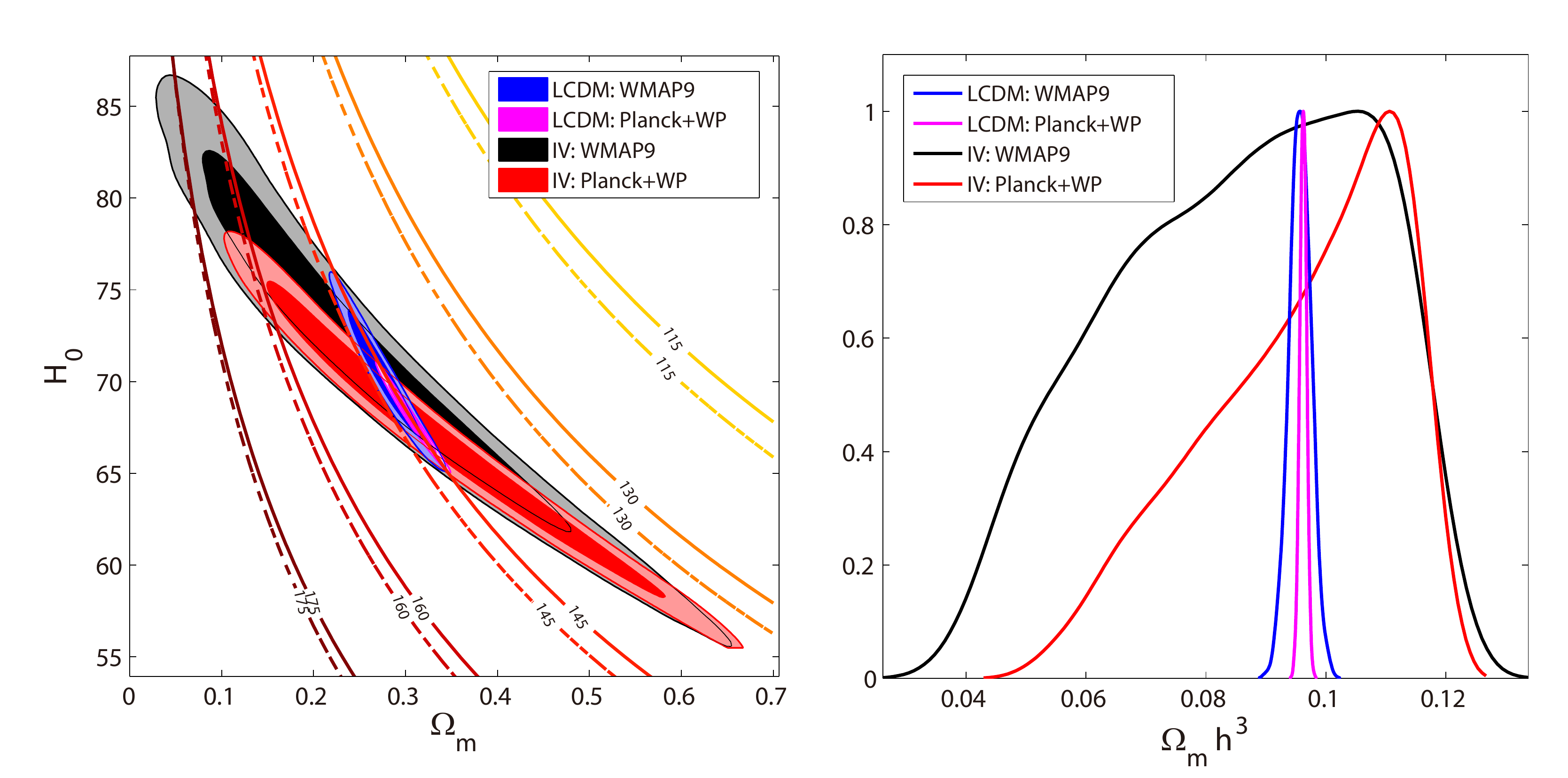}  \\
  \caption{Left panel: the 2D contours of $\Omega_{\rm \tc{m}}-H_0$ at the 68\% C.L. and 95\% C.L. and the contour curves (solid lines: $\Lambda$CDM and dashed lines: interacting vacuum with $\alpha=0.3$) of constant sound horizon at the time of last-scattering, $r_s(z_{\ast})$, in units of Mpc; Right panel: the 1D marginalized distributions of the parameter, $\Omega_{\rm \tc{m}}h^3$.}\label{fig:h0-omegam-rs}
\end{figure*}

\tc{In this section, we shall present the latest observational constraint on the interacting vacuum model using various kinds of data including CMB, BAO, SN Ia and RSD and their combinations. The $\Lambda$CDM model is also constrained using the same \mv{CMB data} for the purpose of comparison.}

We start from the CMB data of WMAP9 and Planck+WP respectively, and present the results in Table \ref{tab:9-pl-only-results} and Figs.~\ref{fig:1D-H0-com} - \ref{fig:h0-omegam-rs}. For the $\Lambda$CDM model, it is apparent that the constraint on $H_0$ using Planck+WP ($H_0 = 68.0\pm1.2$ km~s$^{-1}$\,Mpc$^{-1}$) is in tension with the HST measurement, $H_0 = 73.8\pm2.4$ km~s$^{-1}$\,Mpc$^{-1}$ , as shown in Table \ref{tab:9-pl-only-results} and in Fig.~\ref{fig:1D-H0-com}. However, in the interacting vacuum model, \ie, when $\alpha$ is allowed to vary, the tension \dw{may be alleviated}. This is because the constraint on $H_0$ \dw{is weakened when we marginalise over different values of $\alpha$, and} the error bar is enlarged from $\pm1.2$ to $\pm5.5$ km~s$^{-1}$ Mpc$^{-1}$ with the central value slightly lowered by 1.5\%. \dw{In particular, positive $\alpha$ accommodates larger $H_0$.} \mvs{The constraint on $H_0$ using WMAP9 in the interacting vacuum model is more consistent with the HST measurement than with lowH value. In the $\Lambda$CDM model, the WMAP9 result for $H_0$ is about $1\,\sigma$ lower than HST value and in better agreement with the lowH prior, see Fig.~\ref{fig:1D-H0-com}. The Planck+WP estimates of $H_0$ in both models agree with the lowH prior.}

\tc{The constraint on $\alpha$ using CMB data alone is rather loose: even Planck+WP cannot distinguish the interacting vacuum models with $|\alpha|\simeq 0.3$ from the $\Lambda$CDM model ($\alpha=0$). This is because of the strong degeneracy between $\alpha$ and $\Omega_{\rm \tc{m}}h^3$, as shown in the left panel of Fig.~\ref{fig:alpha-omegah3}. This degeneracy can be understood using left panel of Fig.~\ref{fig:h0-omegam-rs}. As shown, the $\alpha=0.3$ model has a \mv{smaller sound horizon (the dashed curves with the same values of sound horizon as those in the $\Lambda$CDM model move to the larger values end with respect to the $\Lambda$CDM model)}, and this change can easily be mimicked by tuning $\Omega_{\rm m}$ and $h$ (hence $\Omega_{\rm \tc{m}}h^3$).}

\tc{Due to this degeneracy, the constraint on other cosmological parameters, especially for $\Omega_{\rm m}$, $H_0$, $\sigma_8$, are largely diluted, which can be seen in the contour plots Figs.~\ref{fig:alpha-omegah3} (right panel) and \ref{fig:h0-omegam-rs} (left), and in the 1D posterior distribution plot Fig.~\ref{fig:h0-omegam-rs} (right).}

\begin{figure*}[!htbp]
\includegraphics[scale=0.55]{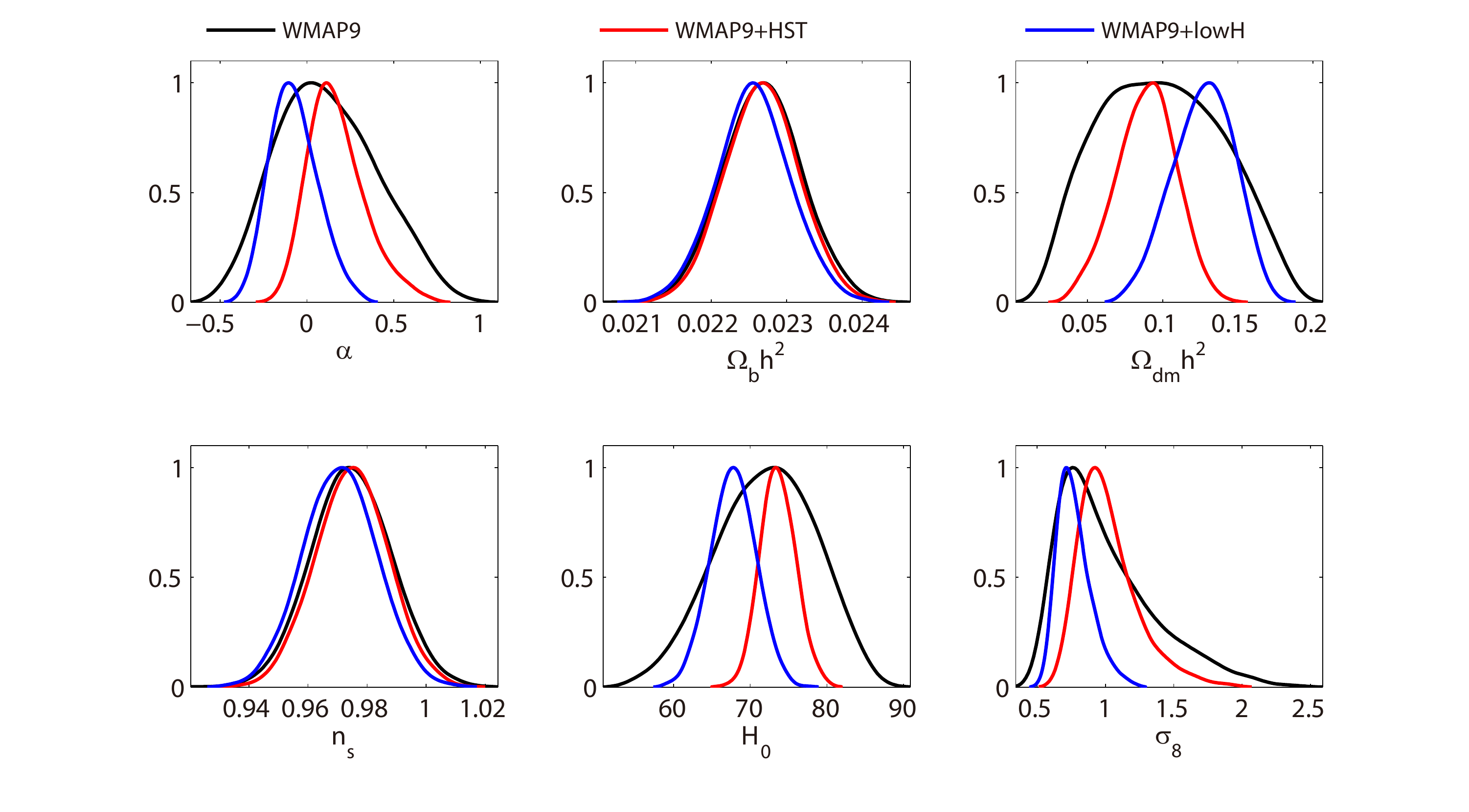}\\
  \caption{The 1D marginalized distributions of some parameters in the interacting vacuum model from the WMAP9 alone and WMAP9 combined with a HST prior and with a lowH prior respectively.}\label{fig:w9-1D-h}
\end{figure*}

\begin{figure*}[!htbp]
\includegraphics[scale=0.55]{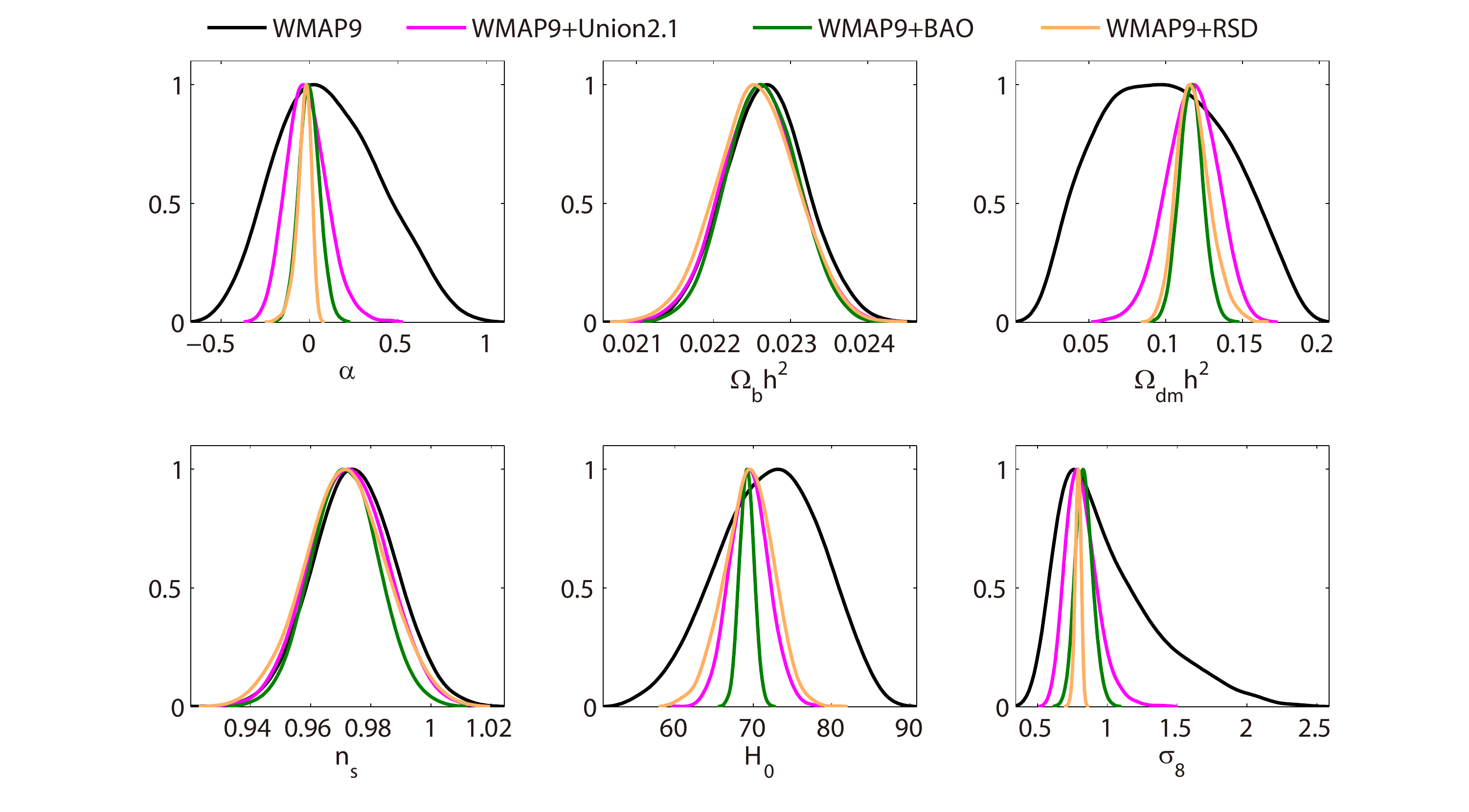}\\
  \caption{The 1D marginalized distributions of some parameters in the interacting vacuum model from the WMAP9 alone and WMAP9 combined with other data (SN Ia, BAO and RSD) respectively.}\label{fig:w9-1D-o}
\end{figure*}

\begin{figure*}[!htbp]
\includegraphics[scale=0.55]{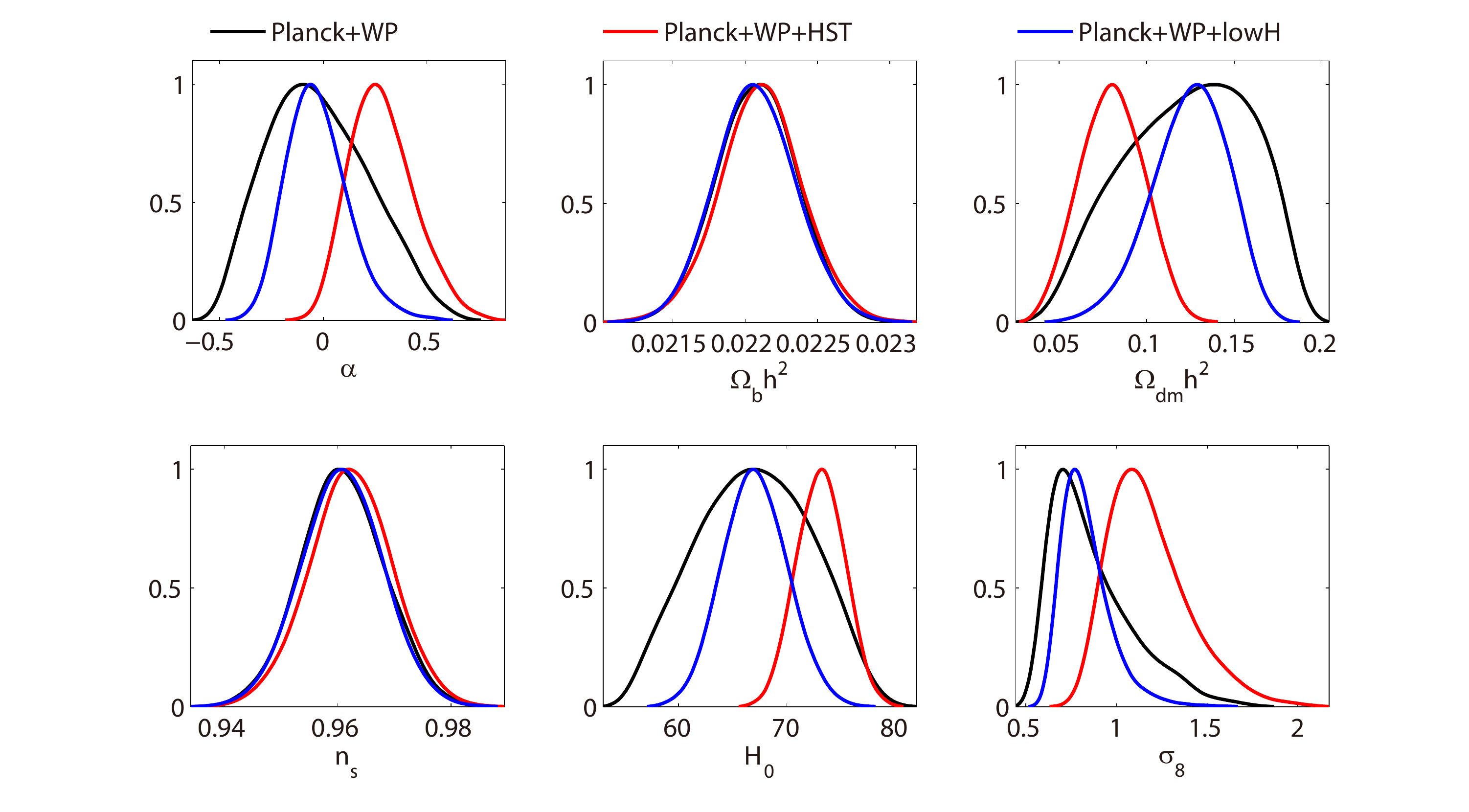}\\
  \caption{The 1D marginalized distributions of some parameters in the interacting vacuum model from the Planck+WP alone and Planck+WP combined with a HST prior and with a lowH prior respectively.}\label{fig:pl-1D-h}
\end{figure*}

\begin{figure*}[!htbp]
\includegraphics[scale=0.55]{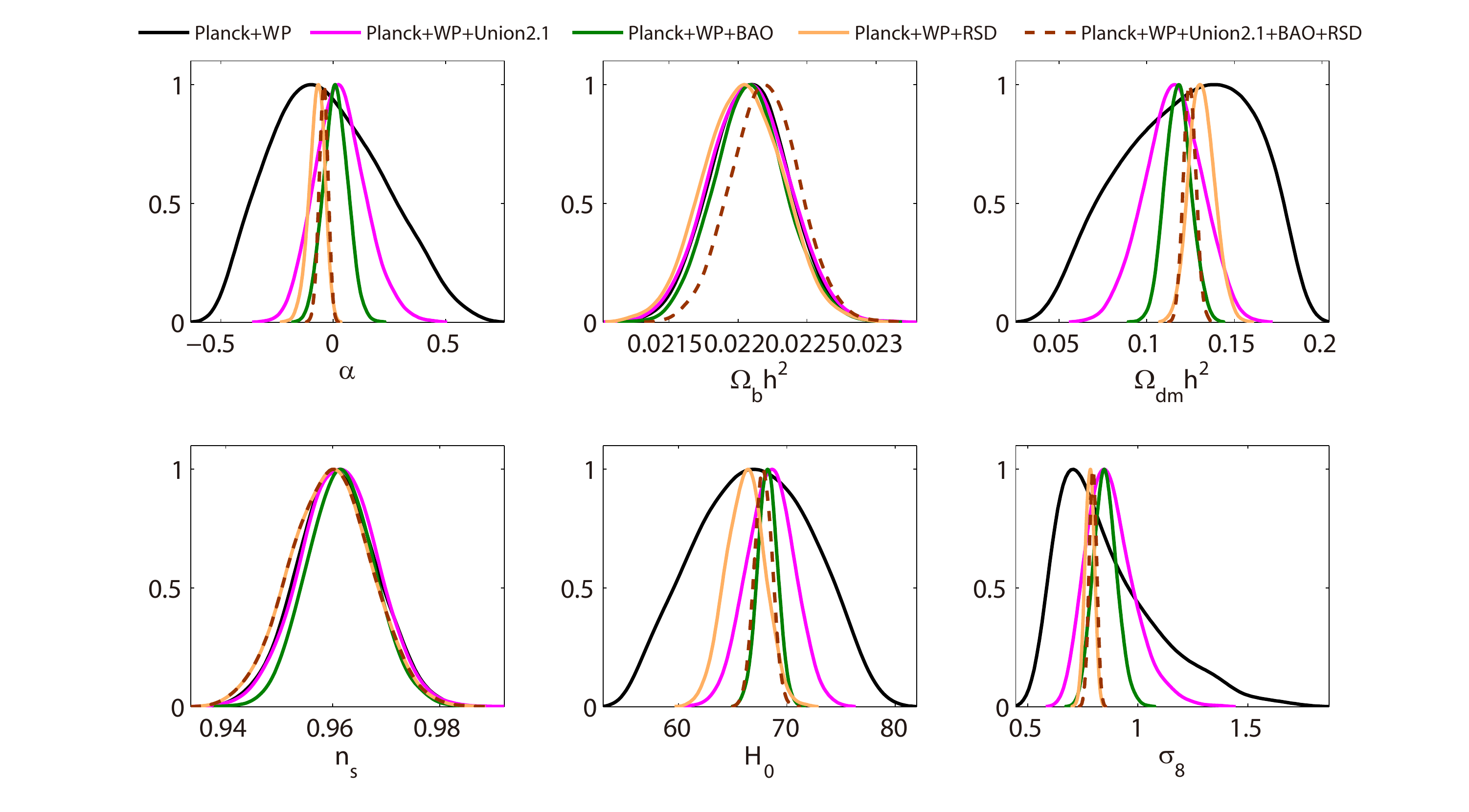}\\
  \caption{The 1D marginalized distributions of some parameters in the interacting vacuum model from the Planck+WP alone and Planck+WP combined with other data (SN Ia, BAO and RSD) respectively, and with these data together.}\label{fig:pl-1D-o}
\end{figure*}

\begin{figure*}[!htbp]
\includegraphics[scale=0.6]{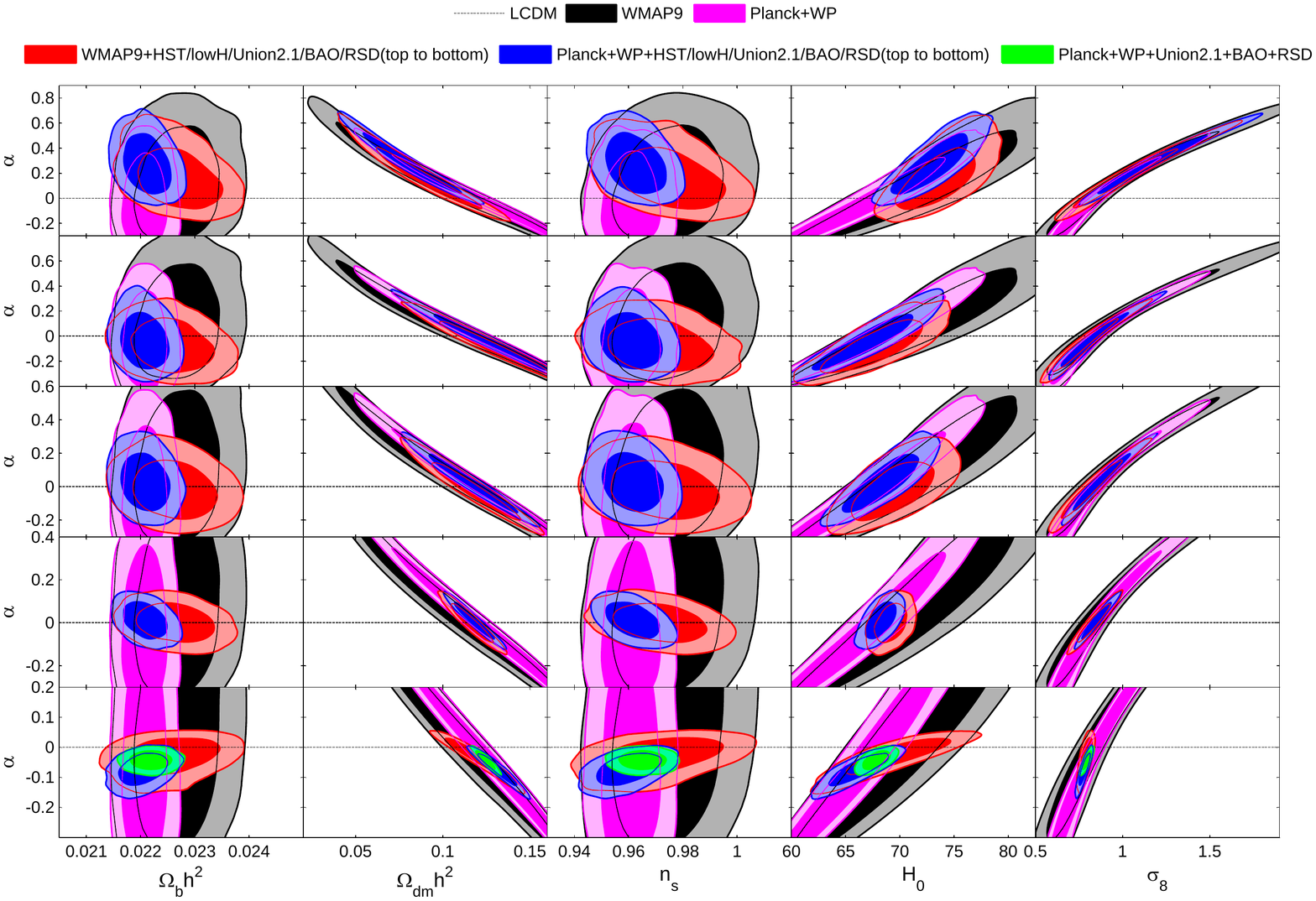}\\
  \caption{The 2D contours with 68\% C.L. and 95\% C.L. between the interacting parameter and some other cosmological parameter in the interacting vacuum model from the CMB data alone and their combinations with other data.}\label{fig:all-2D}
\end{figure*}

\tc{The degeneracy between $\alpha$ and $\Omega_{\rm \tc{m}} h^3$ can be broken by combining with additional datasets including \mvs{the $H_0$ prior (HST or lowH)}, SN Ia (Union2.1), BAO, and RSD. The results using these multiple probes are shown in Tables \ref{tab:9-pl-each-results}, \ref{tab:final-results} and in Figs.~\ref{fig:w9-1D-h}, \ref{fig:w9-1D-o}, \ref{fig:pl-1D-h}, \ref{fig:pl-1D-o} (1D posterior likelihood distributions), and \ref{fig:all-2D} (2D contours). We firstly combine CMB data (WMAP9 and Planck respectively) with other data including, HST, \mvs{lowH}, SN Ia, BAO and RSD, one at a time, then combine Planck with all other datasets. The key points from these plots / tables include,}
\tc{
\begin{enumerate}
\item Figs.~\ref{fig:w9-1D-h} and \ref{fig:pl-1D-h}: since WMAP9 is consistent with the HST measurement \dw{of $H_0$}, adding the HST prior does not change the mean value of $H_0$ much, but greatly \dw{shrinks the allowed range for} $H_0$. However, Planck+WP is in tension with the HST \dw{measurement of} $H_0$. Thus combining Planck+WP with a HST prior not only \dw{shrinks the error bar, but also shifts} the central value of $H_0$. \mvs{By comparison with the case of a lowH prior, the results are reversed. The $H_0$ derived from Planck+WP and the lowH prior agree, so combining these two data hardly affects the mean value of $H_0$, but tightens the uncertainties of $H_0$. WMAP9 favors a high mean value, so adding the lowH prior to WMAP9 causes the change of the mean value of $H_0$ and the improvements of the errors of $H_0$;}
\item Table \ref{tab:9-pl-each-results}: \mvs{to conclude HST prior/lowH prior comparison, we find that the combinations of HST prior and CMB (WMAP9 or Planck+WP) favor high $H_0$ and positive $\alpha$, whereas combining lowH with WMAP9 and with Planck+WP respectively give lower $H_0$ and smaller $\alpha$, even negative mean values.}
\item Figs.~\ref{fig:w9-1D-o} and \ref{fig:pl-1D-o}: \mv{since BAO provides a measurement of expansion history of the Universe, combining CMB with BAO gives the tightest constraints on $H_0$. RSD data provides a measurement of the growth of structure and thus can greatly improve the bounds on $\sigma_8$};
\item Fig.~\ref{fig:all-2D}: $\alpha$ \dw{is positively correlated with $H_0$ and $\sigma_8$, and anti-correlated with $\Omega_{\rm dm}h^2$ for any data sets};
\item Figs.~\ref{fig:w9-1D-o} and \ref{fig:pl-1D-o}: RSD is \mv{the} most powerful data to constrain $\alpha$ in combination with CMB data, thus the allowed range for $\alpha$ is minimised when combining RSD with CMB data;
\item Table \ref{tab:9-pl-each-results}: Planck+WP combined with HST or RSD, \dw{gives a preference for $\alpha>0$ at 1.56 $\sigma$ or $\alpha<0$ at 1.85 $\sigma$ respectively. By comparison, the $\Lambda$CDM model lies within the $1\,\sigma$ region using constraints from other datasets};
\item Table \ref{tab:final-results}: combining Planck+WP with SN Ia, BAO and RSD, we measure $\alpha$ to be,
\begin{equation}\label{eq:alpha}
  \alpha= -0.043_{-0.020-        0.040}^{+        0.019+        0.037}
\end{equation}
\dw{where sub- and super-scripts denote $1$ and $2\sigma$ constraints. Thus we find evidence for negative $\alpha$ at the $2\,\sigma$ level}. This is largely due to the inclusion of RSD data; a negative $\alpha$ means a lower growth rate than that in the $\Lambda$CDM model, which is what we have seen in Fig.~\ref{fig:f8-sigma8} (upper panel). An interacting vacuum model with a negative $\alpha$ provides one possible solution to the problem of the tension between RSD and \dw{CMB} measurements in the $\Lambda$CDM model \dw{\cite{Macaulay:2013swa}, but it cannot relieve tension with HST measurements of $H_0$ at the same time.} \mvs{Negative $\alpha$ allows lower $H_0$, being in good agreement with lowH value;}
\item Table \ref{tab:final-results}: \mv{comparing our results with previous results in Ref.~\cite{Wang:2013Jan}, we find that the error bars on the interaction parameter, $\alpha$, are improved by nearly an order of magnitude from $|0.1|$ to $|0.02|$.}
\end{enumerate}
}

\begin{table*}
\begin{center}
\begin{tabular}{|c|c|c|c|c|c|c|c|c|c|}
\hline \hline
 Data & $\alpha$ & $\Omega_bh^2$& $\Omega_{\rm \tc{dm}}h^2$& $n_s$ &$H_0$ &$\sigma_8$& C.L.\\ \hline
 WMAP9+HST &$       ~~0.173_{-        0.196}^{+        0.123}$&
            $        0.0227_{-        0.0005}^{+        0.0005}$&
            $        0.0890_{-        0.0192}^{+        0.0213}$&
            $        0.975_{-        0.012}^{+        0.012}$&
            $        73.6_{-        2.4}^{+        2.3}$&
            $        1.024_{-        0.263}^{+        0.123}$&
            $0.88\,\sigma$\\
  WMAP9+lowH &$       -0.074_{-        0.160}^{+        0.120}$&
            $        0.0226_{-        0.0005}^{+        0.0005}$&
            $        0.1270_{-        0.0192}^{+        0.0235}$&
            $        0.971_{-        0.012}^{+        0.012}$&
            $        67.8_{-        2.9}^{+        2.9}$&
            $        0.774_{-        0.152}^{+        0.082}$&
            $0.46\,\sigma$\\
 WMAP9+Union2.1&$       -0.011_{-        0.127}^{+        0.101}$&
                $        0.0226_{-        0.0005}^{+        0.0005}$&
                $        0.1167_{-        0.0157}^{+        0.0179}$&
                $        0.973_{-        0.013}^{+        0.013}$&
                $        69.5_{-        2.4}^{+        2.4}$&
                $        0.823_{-        0.136}^{+        0.083}$&
                $ 0.08\,\sigma$\\
  WMAP9+BAO  &$       -0.001_{-        0.057}^{+        0.057}$ &
              $       0.0226_{-        0.0004}^{+        0.0004}$&
              $       0.1164_{-        0.0076}^{+        0.0075}$&
              $        0.972_{-        0.011}^{+        0.011}$&
              $        69.2_{-        0.9}^{+        0.9}$&
              $        0.834_{-        0.068}^{+        0.057}$&
              $0.01\,\sigma$ \\
 WMAP9+RSD   &$       -0.032_{-        0.028}^{+        0.049}$&
              $        0.0226_{-        0.0005}^{+        0.0005}$&
              $        0.1183_{-        0.0123}^{+        0.0088}$&
              $        0.972_{-        0.013}^{+        0.013}$&
              $        69.4_{-        2.9}^{+        3.3}$&
              $        0.791_{-        0.019}^{+        0.023}$&
              $0.65\,\sigma$\\\hline
 Planck+WP+HST &$        ~~0.286_{-        0.183}^{+        0.130}$&
                $        0.0221_{-        0.0003}^{+        0.0003}$&
                $        0.0801_{-        0.0183}^{+        0.0183}$&
                $        0.962_{-        0.007}^{+        0.007}$&
                $        73.1_{-        2.2}^{+        2.2}$&
                $        1.179_{-        0.270}^{+        0.140}$&
                $1.56\,\sigma$\\
 Planck+WP+lowH &$       -0.027_{-        0.169}^{+        0.122}$&
                $        0.0221_{-        0.0003}^{+        0.0003}$&
                $        0.1245_{-        0.0188}^{+        0.0249}$&
                $        0.961_{-        0.007}^{+        0.007}$&
                $        67.1_{-        2.9}^{+        2.9}$&
                $        0.834_{-        0.162}^{+        0.076}$&
                $0.16\,\sigma$\\
 Planck+WP+Union2.1&$      ~~0.033_{-        0.123}^{+        0.102}$&
                    $        0.0221_{-        0.0003}^{+        0.0003}$&
                    $        0.1150_{-        0.0162}^{+        0.0163}$&
                    $        0.961_{-        0.007}^{+        0.007}$&
                    $       68.5_{-        2.2}^{+        2.1}$&
                    $        0.880_{-        0.132}^{+        0.080}$&
                    $ 0.26\,\sigma$\\
 Planck+WP+BAO  &$        ~~0.011_{-        0.054}^{+        0.053}$&
                 $        0.0221_{-        0.0003}^{+        0.0002}$&
                 $        0.1178_{-        0.0077}^{+        0.0070}$&
                 $        0.962_{-        0.006}^{+        0.006}$&
                 $       68.2_{-        0.9}^{+        0.9}$&
                 $        0.852_{-        0.057}^{+        0.050}$&
                 $0.20\,\sigma$ \\
 Planck+WP+RSD&$       -0.074_{-        0.030}^{+        0.040}$&
               $        0.0220_{-        0.0003}^{+        0.0003}$&
               $        0.1315_{-        0.0079}^{+        0.0067}$&
               $        0.960_{-        0.008}^{+        0.007}$&
               $       66.1_{-        1.8}^{+        1.7}$&
               $        0.781_{-        0.021}^{+        0.021}$&
               $ 1.85\,\sigma$ \\
\hline \hline
\end{tabular}
\caption{Mean values of parameters with $1\,\sigma$ limits for the interacting vacuum model using different data combinations. \mv{The last column is the confidence level at which the mean value of the interaction parameter, $\alpha$, departs from zero (the $\Lambda$CDM model)}.}\label{tab:9-pl-each-results}
\end{center}
\end{table*}

\begin{table}
\begin{center}
\begin{tabular}{|c|c|c|}
\hline\hline Combined data&\multicolumn{2}{|c|}{Planck+WP+Union2.1+BAO+RSD}\\ \hline
Parameters&Mean values with $1\,\sigma, 2\,\sigma$ errors& Best fit\\ \hline
$\Omega_b h^2$ & $        0.0222_{-        0.0002-        0.0005}^{+        0.0002+        0.0005}$ & $0.0222$\\
$\Omega_{\rm \tc{dm}} h^2$ & $        0.1245_{-        0.0034-        0.0067}^{+        0.0035+        0.0070}$ &$0.1225$\\
$\Theta_S$ &$        1.0412_{-        0.0006-        0.0011}^{+        0.0006+        0.0011}$ &$1.0416$\\
$\tau$ & $        0.0884_{-        0.0137-        0.0241}^{+        0.0124+        0.0258}$ &$0.0925$\\
$n_s$ & $        0.965_{-        0.005-        0.011}^{+        0.006+        0.011}$ &$0.969$\\
${\rm{ln}}(10^{10} A_s)$ &  $        3.082_{-        0.026-        0.047}^{+        0.024+        0.050}$ &$3.088$\\\hline
$\alpha$ & $       -0.043_{-        0.020-        0.040}^{+        0.019+        0.037}$ &$-0.036$\\ \hline
$\Omega_V$ & $        0.681_{-        0.014-        0.029}^{+        0.015+        0.027}$&$0.691$\\
$\Omega_{\rm \tc{m}}$ & $        0.319_{-        0.015-        0.027}^{+        0.014+        0.029}$ &$0.309$\\
$\sigma_8$ &  $        0.796_{-        0.016-        0.033}^{+        0.017+        0.032}$ &$0.801$\\
$H_0$ & $       67.8_{-        0.8-        1.6}^{+        0.8+        1.6}$ &$68.4$\\
$\Omega_{\rm \tc{m}} h^3$ & $        0.0995_{-        0.0016-        0.0031}^{+        0.0016+        0.0032}$ &$0.0990$\\
\hline\hline
\end{tabular}
\caption{Mean values with $1\,\sigma, 2\,\sigma$ limits and best fit values of parameters for the interacting vacuum model using the combination of Planck+WP+Union2.1+BAO+RSD.}\label{tab:final-results}
\end{center}
\end{table}

\section{Conclusions and Discussions} \label{summary}

An interacting vacuum model provides an interesting alternative dark energy model in which to interpret the cosmological parameter constraints coming from the latest CMB data in combination with other data sets. Unlike other dark energy models such as non-vacuum fluid or scalar field \tc{models}, there are no additional degrees of freedom if the vacuum energy transfers energy-momentum to or from existing matter fields.

In this paper we have considered \tc{a} particular example of an interacting vacuum cosmology, where the interaction is characterised by a single dimensionless parameter, $\alpha$, which reproduces the background dynamics of a GCG cosmology. However, we have considered a decomposed GCG model where the energy-momentum transfer, from dark matter to vacuum, is always proportional to the matter 4-velocity. As a result the dark matter particles follow geodesics \cite{Wands:2012Mar}, and in the limit of a vanishing interaction parameter, $\alpha\to0$, we recover the $\Lambda$CDM cosmology.

We have used the latest observational data to test the model parameters, and in particular the interaction parameter, $\alpha$, against CMB data alone (WMAP9 or Planck+WP) and various combinations with other data, including the direct measurement of $H_0$ from HST, \mvs{the relatively low $H_0$ value measured from 6dF Galaxy Survey,} the Union2.1 supernova compilation, baryon acoustic oscillations and redshift-space distortions.

In particular possible tension between Planck+WP constraints on $H_0$ and HST measurements is investigated in the interacting vacuum model. Using the WMAP9 alone, we obtain a value of Hubble constant, $H_0=72.1_{-6.2}^{+7.2}$ km\,s$^{-1}$\,Mpc$^{-1}$ (68\% C.L.), which is entirely consistent with the direct measurement of $H_0$ from HST. On the other hand, Planck+WP require $H_0= 67.0_{-5.5}^{+5.5}$ km\,s$^{-1}$\,Mpc$^{-1}$ (68\% C.L.). The low mean value from Planck+WP is discrepant with the HST measurement to $H_0$. However, there exists overlap between the marginalized distribution of $H_0$ from Planck+WP and the values of $H_0$ with $1\,\sigma$ errors from HST measurement. \mvs{Compared with another $H_0$ measurement from 6dF Galaxy Survey, it is found that the $H_0$ result from Planck+WP is in better agreement than that from WMAP9.} \tc{The constraint using CMB alone on the interacting vacuum model interaction parameter is too weak to be distinguished from the $\Lambda$CDM model.}

Next, we combined CMB data with other data including the HST prior on $H_0$, \mvs{another low $H_0$ prior,} Union2.1 SN Ia, BAO or RSD. The combined data-sets can break degeneracies, yielding tighter constraints. The constraints on the interaction parameter from the combinations of WMAP9 and other data show that the interacting vacuum model is indistinguishable from the $\Lambda$CDM model within $1\,\sigma$ region. For the predictions of Hubble constant in the interacting vacuum model from different data, we find that the WMAP9 alone and WMAP9+HST favour high values of $H_0$, consistent with the HST prior.

Using Planck+WP in combination with the HST prior on $H_0$ would favour a positive interaction, $\alpha>0$. However constraints on the Hubble constant in the interacting vacuum model using Planck+WP, \mvs{Planck+WP+lowH,} Planck+WP+Union2.1, Planck+WP+BAO and Planck+WP+RSD all yield low values for $H_0$, indicating a tension between Planck+WP and HST measurements of $H_0$. RSD are particularly sensitive to the interaction parameter and Planck+WP+RSD favour a negative interaction \tc{at more than 1.8 $\sigma$ level}.

Finally, based on the above discussions about the consistency of Planck+WP and other data-sets, we use the combined data of Planck+WP+Union2.1+BAO+RSD to constrain the interacting vacuum model. A strong constraint on the interacting vacuum parameter is obtained, $\alpha= -0.043_{-0.020- 0.040}^{+0.019+0.037}$. We conclude that there is \tc{a hint} for a negative energy transfer $\alpha<0$ in the interacting vacuum model at 95\% confidence level. \tc{This model provides a possible solution to the problem of tension between the RSD and other measurements in the $\Lambda$CDM model. }

It would be interesting to investigate further the Bayesian evidence for departures from $\Lambda$CDM using different criteria \cite{Liddle:2004nh} both in this particular decomposed GCG model and in more general interacting vacuum energy models. Negative $\alpha$ implies a slower growth rate for linear density perturbations and thus a lower value for $\sigma_8$. Thus one might also expect lower cluster number counts than predicted in $\Lambda$CDM \cite{Ade:2013lmv}. However halo collapse is a non-linear process and we have not yet studied non-linear collapse in this model. Our assumption that the energy-momentum transfer is proportional to the matter 4-velocity implies that the 4-velocity is proportional to the gradient of the vacuum energy, $u_\mu\propto \nabla_\mu V$, and thus irrotational. Recently Sawicki et al \cite{Sawicki:2013wja} have argued that non-linear collapse in unified dark matter models with irrotational flow will lead to the formation of central black holes rather than virtualised halos. \dw{Either the assumption of a irrotational flow must break down on some scale, or we would require only some fraction of the dark matter to interact with the vacuum energy in this way (and hence be irrotational). In this case the small value required for the interaction parameter $\alpha$ might represent the small fraction of dark matter which collapses into central (supermassive) black hole during halo collapse}. Investigation of this goes beyond the study of linear perturbation theory used in this paper and we leave this as an interesting open issue for future work.

\dw{Note added: We have not included the latest BICEP2 results \cite{Ade:2014xna}  which appeared while this paper was in preparation. These remarkable results show evidence for primordial gravitational waves at the time of CMB last scattering. If confirmed this implies there will be an additional contribution from gravitational waves to the CMB temperature power spectrum at low $\ell$ which appears to be in tension with the minimal $\Lambda$CDM model. It seems unlikely that an interacting vacuum model alone, whose main effect is to change the relation between CMB anisotropies and structure formation at late times, can resolve this apparent tension at low $\ell$. It will be interesting to study this in a broader class of models including interacting vacuum energy.}

\acknowledgements{We thank Marco Bruni and Rob Crittenden for helpful comments. \tc{YW is supported by \mvs{the China Postdoctoral Science Foundation grant No. 2014M550091, by the Young Researcher Grant of} National Astronomical Observatories, Chinese Academy of Sciences, and by University of Portsmouth. GBZ is supported by the $1000$ Young Talents program in China, by the 973 Program grant No. 2013CB837900, NSFC grant No. 11261140641, and CAS grant No. KJZD-EW-T01, by National Astronomical Observatories, Chinese Academy of Sciences, and by University of Portsmouth.} YW and GBZ are supported by the Strategic Priority Research Program ``The Emergence of Cosmological Structures" of the Chinese Academy of Sciences, Grant No. XDB09000000. DW is supported by STFC grants ST/K00090/1 and ST/L005573/1. LX is supported in part by NSFC under the Grants No.\,11275035 and ``the Fundamental Research Funds for the Central Universities" under the Grants No.\,DUT13LK01. All the numerical computations were \tc{performed} on the Sciama High Performance Compute cluster, supported by the SEPNet and University of Portsmouth.}

\end{document}